\documentclass[aps,prl,twocolumn,showpacs,superscriptaddress,groupedaddress,amsmath,amssymb]{revtex4-1}  % for review and submission
\usepackage{graphicx}  % needed for figures
\usepackage[caption=false]{subfig}
\DeclareMathOperator{\Tr}{Tr}
\newcommand{\Pplus}{\ensuremath{\mathcal{P}_{+}}}
\newcommand{\Pstar}{\ensuremath{\mathcal{P}^*}}
\def\ket #1{\vert #1\rangle}

\newcommand{\Jam}{Jamio\l kowski }
\usepackage{longtable}
\usepackage{bbold}
\usepackage{hyperref}
\usepackage{url}
\hypersetup{colorlinks,breaklinks,
citecolor=[rgb]{0.0,0.5,0.5},
    urlcolor=[rgb]{0.0,0.5,0.5},
    linkcolor=[rgb]{0.0,0.5,0.5}}

\def\id{\mbox{\small 1} \!\! \mbox{1}}
\newcommand{\tr}{\ensuremath{\mathrm{\Tr}}}

\begin{document}

\title{Application of a resource theory for magic states to fault-tolerant quantum computing}
\author{Mark Howard}
\email{m.howard@sheffield.ac.uk}
\affiliation{Department of Physics and Astronomy, University of Sheffield,
Sheffield, UK
}
\author{Earl Campbell}
\email{earltcampbell@gmail.com}
\affiliation{Department of Physics and Astronomy, University of Sheffield,
Sheffield, UK
}

\begin{abstract}
Motivated by their necessity for most fault-tolerant quantum computation schemes, we formulate a resource theory for magic states.
We first show that robustness of magic is a well-behaved magic monotone that operationally quantifies the classical simulation overhead for a Gottesman-Knill type scheme using ancillary magic states.
Our framework subsequently finds immediate application in the task of synthesizing non-Clifford gates using magic states. When magic states are interspersed with Clifford gates, Pauli measurements and stabilizer ancillas---the most general synthesis scenario---then the class of synthesizable unitaries is hard to characterize. Our techniques can place non-trivial lower bounds on the number of magic states required for implementing a given target unitary. Guided by these results we have found new and optimal examples of such synthesis. 
\end{abstract}

%\pacs{}
\maketitle

Quantum resource theories attempt to capture what is quintessentially quantum in a piece of technology. For example, entanglement is the relevant resource for quantum cryptography and communication. The resource framework for entanglement finds practical application in bounding the efficiency of entanglement distillation protocols. An abundance of other resource theories have been related to various aspects of quantum theory \cite{brandao_reversible_2015,coecke_mathematical_2016,grudka_quantifying_2014,Horodecki:2013,napoli_robustness_2016,stahlke_quantum_2014,veitch_resource_2014,vidal_robustness_1999}. Once a quantum computer is made fault-tolerant, some computational operations become relatively easy, and some more difficult, leading to a natural resource picture called the magic state model~\cite{bravyi_universal_2005,knill_quantum_2005} (although alternative routes to fault-tolerant universality exist \cite{campbell_steep_2016}). Preparation of stabilizer states and implementation of Clifford unitaries and Pauli measurements constitute free resources. Difficult operations include preparation of  magic states, a supply of which is necessary in order to promote the easier operations to a universal gate set.   With only free resources, the computation can be efficiently classically simulated, whereas with a liberal supply of pure magic states, universal quantum computation is unlocked.    For qudit ($d$-level) quantum computers with odd $d$, a resource theory of magic (or equivalently contextuality with respect to stabilizer measurements \cite{howard_contextuality_2014,delfosse_equivalence_2016}) has been developed~\cite{veitch_negative_2012,mari_positive_2012,veitch_resource_2014}. This relies on a well-behaved discrete Wigner function \cite{gross_hudsons_2006}, which in turn relies on quirks of odd dimensional Hilbert space. Here we address the most practically important case by quantifying the magic for multiqubit systems, relating this resource measure to simulation complexity and applying the resource theory to the practical problem of gate synthesis.

The canonical magic state is $\ket{H}=(\ket{0}+e^{i\pi/4}\ket{1})/\sqrt{2}$, which enables application of a single-qubit unitary $T=\mathrm{diag}(1,e^{i\pi/4})$ \cite{bravyi_universal_2005,knill_quantum_2005}.  A circuit composed of elements from the Clifford+$T$ gate set acting on the standard computational basis input suffices for universal quantum computation.  Such a circuit can be classically simulated, but in a time that scales exponentially in the number of $T$ gates~\cite{aaronson_improved_2004}.  Faster simulation algorithms were recently discovered that relate the simulation complexity to the stabilizer rank~\cite{bravyi_improved_2016,bravyi_trading_2016,garcia_geometry_2014}, a measure of magic for pure states. Such techniques do not naturally adapt to mixed magic states, and stabilizer rank is qualitatively very different to the magic measure we establish here.  For quantum computations using qudits with odd dimension, the discrete Wigner function provides a quasiprobabilitiy distribution and Pashayan et al.~\cite{pashayan_estimating_2015} showed that the negativity quantifies the simulation complexity.  Here we provide a general simulation scheme, which can be naturally applied to mixed-state qubit quantum computations using any kind of ancillary magic state (e.g., a multi-qubit magic state enabling a Toffoli gate). Furthermore, for many problems our approach is competitive with comparable schemes based on stabilizer rank \cite{bravyi_improved_2016,bravyi_trading_2016}. 

% By advancing simulation techniques, we push up the scale of a quantum computer required to outperformance their best classical counterparts.

Due to the high price assigned to $\ket{H}$ states and hence $T$ gates, it behooves us to find Clifford$+T$ circuit implementations of quantum algorithms that are parsimonious in their use of $T$ gates. The topic of circuit synthesis has made tremendous progress in recent years~\cite{amy_polynomial-time_2014,amy_t-count_2016,gosset_algorithm_2014,bocharov_efficient_2013,duclos-cianci_distillation_2013,paetznick_repeat-until-success:_2014,ross_optimal_2014,selinger_quantum_2013,wiebe_quantum_2016}, compared with  Solovay-Kitaev type constructions that were for a long time the standard benchmark. Developments include identifying special algebraic forms for all gates that can be unitarily synthesized over the Clifford$+T$ gate set~\cite{gosset_algorithm_2014} or over the smaller CNOT+$T$ gate set~\cite{amy_polynomial-time_2014,amy_t-count_2016}. However, circuit synthesis need not be a purely unitary process, and more generally may be aided by ancillary stabilizer states, measurements and classical feed-forward.  There are hints that general synthesis can be significantly more powerful~\cite{duclos-cianci_distillation_2013,jones_low-overhead_2013,paetznick_repeat-until-success:_2014,wiebe_quantum_2016}, though the paradigm is not well understood. Our resource framework helps with this problem by establishing non-trivial lower bounds on the number of $\ket{H}$ states, or equivalently $T$ gates, required for the general synthesis scenario.  This allows us to identify several circuits as optimal. Such resource-theoretic tools work for any kind of state, not just $\ket{H}$, but they are particularly well-motivated for magic states from the third level of the Clifford hierarchy e.g., Toffoli resource states.  We note how general synthesis has a curious relationship to Clifford equivalence of magic states. From this vantage point, we discover several new examples of general synthesis protocols with resource savings over previous unitary synthesis methods. 

 The Supplementary Materials (text, which includes Refs.~\cite{andersson_states_2015,reichardt_quantum_2009,campbell_catalysis_2011,harrow_robustness_2003,van_dam_noise_2011}, and files) contain numerous additional computations. Results include identifying the most robust states, most robust gates and classification of all three and four qubit diagonal gates from the third level of the Clifford hierarchy. Intriguingly, one maximally robust state is the Hoggar \cite{hoggar_64_1998} fiducial state.
%for a symmetric, informationally complete positive operator valued measure.
% (SIC-POVM) covariant under the 3-qubit Pauli group.

% Our discussion highlights some similarities and differences with recent work by other authors, before discussing some limitations as well as interesting open questions raised by our work.

\noindent\textit{Robustness of Magic.---}Vidal and Tarrach~\cite{vidal_robustness_1999} showed that the amount of separable noise that makes an entangled state become separable is an entanglement monotone, which they called robustness. The basic principle can been adapted for use in other resource theories with a set of free states.  Denoting $\mathcal{S}_n=\{ \sigma_i \}$ as the set of pure $n$-qubit stabilizer states, we define the robustness of magic as
\begin{align}
\text{(RoM)}\qquad \mathcal{R}(\rho)=&\min_{x} \left\{ \sum_i | x_i | ;  \rho = \sum_i x_i \sigma_i \right\} \,. \label{eqn:RobustGeom}
\end{align}
Decompositions of the form $\sum_i x_i \sigma_i$ are called stabilizer pseudomixtures.  We have $\sum_{i}x_i=1$, but $x_i$ may be negative and so they provide a quasiprobability distribution.  The optimization in \eqref{eqn:RobustGeom} can be rewritten in terms of a linear system as 
\begin{align}
\mathcal{R}(\rho)=&\min ||x||_1 \ \text{subject to} \ Ax=b\,, \label{eqn:RobustLP}
\end{align}
where $||x||_1=\sum_i |x_i|$, $b_i=\Tr(P_i\rho)$ and $A_{j,i}=\Tr(P_j\sigma_i)$ where $P_j$ is the $j$th Pauli operator. For example, consider the single-qubit magic state $\ket{H}=(\ket{0}+e^{i\pi/4}\ket{1})/\sqrt{2}$, then in the Pauli operator basis
\begin{align*}
A&=\begin{array}{c}
 \langle \id \rangle \\
 \langle X \rangle \\
 \langle Y \rangle \\
 \langle Z \rangle \\
\end{array}\left(
\begin{array}{cccccc}
 1 & 1 & 1 & 1  & 1 & 1  \\
 1 & -1 & 0 & 0 & 0 & 0  \\
 0 & 0 & 1 & -1 & 0 & 0  \\
 0 & 0 & 0 & 0  & 1 & -1 \\
\end{array}
\right),\quad b=\left(
\begin{array}{c}
1 \\
\frac{1}{\sqrt{2}} \\
\frac{1}{\sqrt{2}} \\
0 \\
\end{array}
\right)\,,
\end{align*}
and the solution of \eqref{eqn:RobustLP} is $x=\left(\sqrt{2},0,1,{1-\sqrt{2}},0,0\right)/2$ implying $\mathcal{R}(\ket{H})=\sqrt{2}$. There are a number of freely available solvers for linear programs~\cite{grant_graph_2008,grant_cvx:_2014}, which are efficient in the size of $A$. From our formulation of the problem  it is clear that $\min_{Ax=b} ||x||_1$ is feasible and  bounded.  Consequently, strong duality holds i.e.,
\begin{align}
\min_{Ax=b} ||x||_1 = \max_{||A^Ty||_\infty\leq 1} -b^Ty\,,  \label{eqn:duality}
\end{align}
and the aforementioned solvers can provide a certificate $y$ of optimality \footnote{This certificate of optimality can be used to obtain a magic witness---an operator whose expectation value with respect to stabilizer states is in the interval $[-1,1]$ and whose expectation with respect to $\rho$ is $\mathcal{R}(\rho)$. These witnesses can be used to derive exact expressions for robustness as in Supplementary Material}. Despite the theoretical efficiency of the linear programming problem, the number of stabilizer states in $\mathcal{P}_n$ scales super-exponentially with $n$, so that $|\mathcal{S}_n|=2^{n}\prod_{j=1}^n(2^j+1)$~\cite{gross_hudsons_2006}.  Practically we are limited to $1 \leq n \leq 5$ qubits. We have made available the corresponding $A$ matrices in Supplementary Material.

Robustness of Magic (RoM) possesses all the desirable qualities of a resource theoretic measure (see Supplementary Material for proofs of the following). For a mixed stabilizer state, we find $x_i>0$ entailing $ \sum_i | x_i | =  \sum_i  x_i =1  $.  For a non-stabilizer state, at least one $x_i$ is negative and then RoM must exceed unity. Therefore RoM is faithful. Crucially, RoM is non-increasing under stabilizer operations, the free set of operations in the resource theory. 
%All stabilizer operations act as $\sigma_i \rightarrow \sum_j p_{i,j} \sigma_j$ where $\sum_j |p_{i,j}|= \sum_j p_{i,j}=1$. This entails $\rho \rightarrow \sum_{i,j} x_i p_{i,j} \sigma_j $ and we find $\sum_{i,j} |x_i p_{i,j}| = \sum_{i}|x_i|$. Stabilizer operations may decrease RoM as this new stabilizer pseudomixture may not fulfil the minimisation.   
Finally, RoM is submultiplicative, \begin{align}
\mathcal{R}(\rho_1 \otimes \rho_2)&\leq\mathcal{R}(\rho_1)\mathcal{R}(\rho_2), \label{eqn:submult}
\end{align}
which follows by using the minimal stabilizer pseudomixtures for $\rho_1$ and $\rho_2$ to construct a not-necessarily-minimal stabilizer pseudomixture for $\rho_1 \otimes \rho_2$.  Useful lower bounds on $\mathcal{R}(\rho_1 \otimes \rho_2)$ can also be obtained (see Supplementary Material).  

Resource theoretic frameworks are commonplace in quantum information theory, but do not always directly lend themselves to an operational meaningful interpretation or to useful applications in solving relevant problems.  In the next section we show how RoM quantifies the exponential simulation overhead for a version of the Gottesman-Knill protocol where non-stabilizer ancillas $\rho$ can be added to an otherwise stabilizer circuit. The subsequent section discusses RoM's application to the task of implementing non-Clifford operations in an economical way.
\begin{figure}
\includegraphics[scale=1]{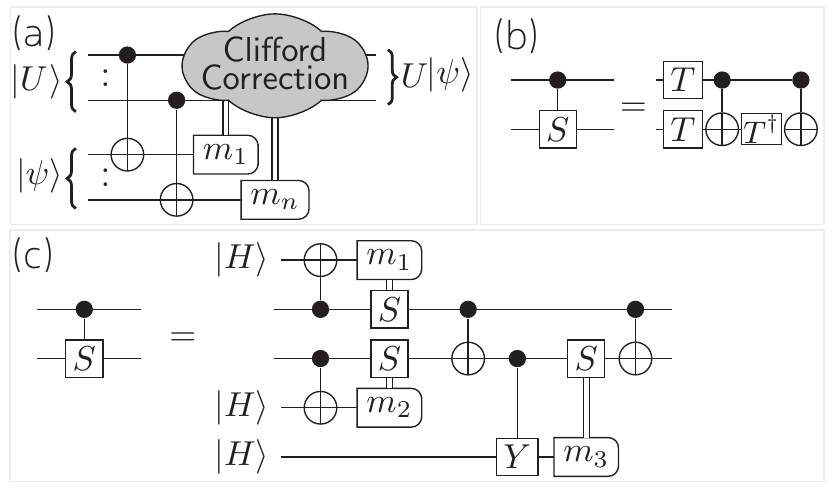}
\caption{\label{fig:CS_half_teleportation} (a) Half-teleportation gadget  for implementing diagonal $U$ in the third level of the Clifford hierarchy. The circuit uses a resource state $\ket{U}=U\ket{+}^{\otimes n}$. (b) Exact synthesis of CS gate using $T$-gates and CNOTS. (c) The CS circuit as a  gadgetized circuit using three $\ket{H}$ magic states. Our techniques show this synthesis is provably optimal.}
\end{figure}

\noindent\textit{Robustness quantifies classical simulation overhead.---}
The Gottesman-Knill (GK) theorem shows that for any stabilizer circuit written as a superoperator $\mathcal{E}$ and pure stabilizer state $\sigma_i$, we can efficiently sample from the outcome of a Pauli measurement on $\mathcal{E}(\sigma_i)$.  By collecting many samples, we can estimate the expectation value to any desired accuracy.  If the input state is a probabilistic ensemble of stabilizer states, the Gottesman-Knill theorem still holds provided we can efficiently sample from the ensemble. A Clifford+$T$ circuit suffering very heavy noise can be simulated in this way \cite{buhrman_new_2006,virmani_classical_2005,plenio_upper_2010}. Here we provide a simulation algorithm for estimating an expectation value  $P_{\rho}:=\mathrm{tr}[P\mathcal{E}(\rho)]$ where $\mathcal{E}$ is a stabilizer operation.  The simulation time cost scales with $\sum_i|x_i| \geq \mathcal{R}(\rho)$ where $x_i$ are the quasiprobabilities used (which may be suboptimal).  First, as in \cite{pashayan_estimating_2015}, we use the quasiprobability distribution $x_i$ to form the probability distribution $p_i=|x_i|/\sum_i|x_i|$.  We sample an $i$ value from this probability distribution and use GK to obtain an eigenvalue $m=\pm 1$ for the Pauli measurement on $\mathcal{E}(\sigma_i)$. Our simulation outputs not $m$, but $M=\mathrm{sign}(x_i)m  \sum_i|x_i|$ where $\mathrm{sign}(x_i)=1$ for $x_i>0$ and $\mathrm{sign}(x_i)=-1$ otherwise.  Notice that each run outputs $\pm \sum_i|x_i|$ and not $\pm 1$, which leads to a larger variance of our random variable.  We repeat this sampling process many times and find the mean value of $M$, which gives an unbiased estimator of $P_{\rho}$. 
%However, the variance of outputs from this quasiprobability simulation with outputs $\pm \sum_i|x_i|$ exceeds the inherent variance of Pauli measurements on $\mathcal{E}(\rho)$.  Therefore,  more samples are needed to obtain a precise estimate.  
The Hoeffding inequalities show that for random variables bounded in the interval $[-\sum_i|x_i|,+\sum_i|x_i|]$,  $N$ samples will estimate the mean to within $\delta$ of the actual mean with probability exceeding $1- \epsilon$ where $\epsilon=2 \exp( - N \delta^2 /2 (\sum_i|x_i|)^2)$.  In other words, the desired accuracy is guaranteed by using
\begin{align}
	N = \frac{2}{\delta^2}(\sum_i|x_i|)^2   \ln \left( \frac{2}{\epsilon} \right)\label{eqn:SampleN}	
\end{align}
samples. Using an optimal stabilizer pseudomixture, the number of samples scales quadratically in the robustness, though the robustness typically scales exponentially in the number of magic states.  For each of these samples, the GK scheme requires a polynomial amount of time provided we know how to efficiently sample from the quasiprobabilitiy distribution.  

\noindent\textit{Nonstabilizer circuits.---}  For any quantum circuit we can find an equivalent gadgetized version~\cite{bravyi_universal_2005,bravyi_improved_2016,bravyi_trading_2016} over the Clifford plus $T$ gate set; all uses of $T$ are replaced with the standard state injection circuit whereby a $\ket{H}$ state is entangled with a data qubit and subsequently measured out (see Fig.~\ref{fig:CS_half_teleportation} for an example). The $T$ gadget is just one example from an infinitely large family of similar gadgets.   All diagonal gates from the third level of the Clifford hierarchy---the set of gates that map Pauli operators to Clifford gates under conjugation---are also suitable for gadgetization. These gates are sufficient to promote Clifford gates to universality, and have the added property that access to the state $\ket{U}=U\ket{+}^{\otimes k}$ allows for deterministic implementation of the gate $U$~\cite{gottesman_demonstrating_1999,zhou_methodology_2000}, as depicted in Fig.~\ref{fig:CS_half_teleportation}.  This family includes important multiqubit gates such as the control-control-Z ($CCZ$), which is Clifford equivalent to the Toffoli, and control-S (CS) where $S=T^2$. Therefore, a quantum circuit $C_1 U_1 C_2 U_2 \ldots U_N C_{N+1}$ where $C_i$ are Clifford is equivalent to a stabilizer circuit consuming the resource $\ket{U}$ where $U:=\otimes U_j$. We remark that diagonal, third-level gates are exactly synthesizable from CNOT and $T$ gates~\cite{amy_polynomial-time_2014,amy_t-count_2016,campbell_unified_2016,campbell_unifying_2016}.

\noindent\textit{Large resource states.---} For large resource states the exact robustness may be difficult to determine.  However,  instead of using the optimal robustness we instead use stabilizer psuedomixtures built up from constant size blocks of qubits, here limited to 5 qubits per block.  For instance, given $t=bm$ copies of the $\ket{H}$ state, we break it into $b$ blocks of $m$-qubits $(\ket{H}^{\otimes m})^{\otimes b}$ and so work with a pseudomixture whose sample complexity scales as $\sum_i|x_i| =  \mathcal{R}( \ket{H}^{\otimes m} )^b=\left[\mathcal{R}( \ket{H}^{\otimes m} )^{\frac{1}{m}}\right]^t $. 

\noindent\textit{Numerical results.---} We performed substantial numerical investigations up to 5 qubit systems, for which there are over two million stabilizer states and over one thousand Pauli operators. For ascending $m\leq 5$ we calculated $\mathcal{R}(\ket{H^{\otimes m}})^{\frac{1}{m}}$ as $\{ 1.414, 1.322, 1.304, 1.301, 1.298 \}$.  The decrease with $m$ shows a strongly submultiplicative behaviour and reduces simulation overheads (though analytic lower bounds derived in Supplementary Materials show going to higher $m$ cannot reduce this value below $1.207$).  Specifically,  $\mathcal{R}(\ket{H^{\otimes 5}})^{\frac{2t}{5}} \approx 1.298^{2t} = 1.685^t$ characterises the complexity of our Clifford plus $t$ $T$ gate simulation. This gives exponential improvement over the method used in~\cite{bravyi_trading_2016} with complexity scaling as $1.9185^t$. A more efficient use of the stabilizer Schmidt decomposition was subsequently established in \cite{bravyi_improved_2016}, leading to complexity scaling as $1.385^t$, although the restriction to pure states as in \cite{bravyi_trading_2016} also holds here. Even better scaling with $t$ can be obtained by using approximate states~\cite{bravyi_improved_2016}, but at the price of $\delta^{-5}$ overhead in the precision $\delta$.  A quantum computation using $z$ $CCZ$ gates, can be implemented using $4z$ $T$ gates~\cite{jones_low-overhead_2013}, implying our simulation complexity scales as $8.067^z$.  However, as discussed earlier, we can use $\ket{CCZ}$ resource states.  We found $\mathcal{R}(\ket{CCZ})=2.555$, and so $z$ $CCZ$s can be simulated with an overhead dominated by $\mathcal{R}(\ket{CCZ})^{2z}=6.531^z$. This gives exponential improvement over using the four-$T$-gate gadgetization. Alternative gadgetizations using $\ket{U}=U\ket{+}$ with third-level diagonal $U$ follow naturally and the simulation overhead is given by $\mathcal{R}(\ket{U})$, see Supplementary Material for many possible examples.

\noindent\textit{Lower bounds on gate synthesis.---} Beyond classical simulation, robustness can help us investigate gate synthesis.  Above, we noted that four $T$ gates can exactly synthesize a $CCZ$ gate.  How can we be sure that a more complicated or clever scheme does not use even fewer $T$ gates, not just for $CCZ$ gates but more generally? This is an important instance where our resource theory of magic applies; a potential resource state $\ket{U}=U\ket{+}$ for a non-Clifford unitary $U$ cannot be made using $t$ $T$ gates if $\mathcal{R}(\ket{U})>\mathcal{R}(\ket{H^{\otimes t}})$. The resource state $\ket{CCZ}$, Clifford-equivalent to a Toffoli resource, has $\mathcal{R}(\ket{CCZ})=2.555$ implying
\begin{align}
\mathcal{R}(\ket{H^{\otimes 3}})<\mathcal{R}(\ket{CCZ})<\mathcal{R}(\ket{H^{\otimes 4}}) .
\end{align}
Therefore, we establish that the four $T$ gate synthesis of $CCZ$ is optimal.  Furthermore, the standard decomposition of $CS$ into 3 $T$ gates (see Fig.~\ref{fig:CS_half_teleportation}) is provably optimal since $\mathcal{R}(\ket{H^{\otimes 2}})<\mathcal{R}(\ket{CS})<\mathcal{R}(\ket{H^{\otimes 3}})$. The numerical results in Supplementary Material allow for the inference of optimality for many more unitaries.

\noindent\textit{Improved gate synthesis.---}  Circuit synthesis can be purely unitary over the Clifford$+T$ gate set, or more generally can make use of stabilizer ancillas and measurement thereby using even fewer $T$ gates. The scale of the potential savings is exemplified by the four $T$ gate realisation of $CCZ$~\cite{jones_low-overhead_2013}, which is assisted by ancillas and measurement.  Purely unitary synthesis of $CCZ$ over Clifford$+T$ is known to need at least seven $T$ gates \cite{gosset_algorithm_2014}.  We shed new light on this phenomena by showing that it emerges from Clifford equivalence of magic states, and give new examples of improved synthesis.  The interesting examples arise when $C\ket{U}=\ket{V}$, and yet unitary synthesis of $U$ uses fewer $T$-gates than $V$.  For these remarkable examples, despite Clifford equivalence of states $\ket{U}$ and $\ket{V}$,  there do not exist Cliffords $C_1$ and $C_2$ such that $C_1 U C_2 = V$. One explicit example, comparable to Jones' construction~\cite{jones_low-overhead_2013}, starts with the $\mathrm{Toff}^*$ gate corresponding to $CCZ_{123}CS_{12}$, which is known to be unitarily synthesizable using four $T$ gates~\cite{selinger_quantum_2013}. Because the ``square-root-of-NOT'' $\sqrt{X}$ is a Clifford gate, we also have that
\begin{align}
    \left(II\sqrt{X}\right) \ket{\mathrm{Toff}^*}=\ket{CCZ}. \label{eqn:ToffCCZ}
\end{align}
Clifford equivalence of magic states provides an alternative proof that $CCZ$ can be performed with four $T$ gates.  

We have found a number of similar examples using the following method (i) Identify $U$ and $V$ with different $T$-count but whose states have the same robustness $\mathcal{R}(\ket{U})=\mathcal{R}(\ket{V})$, (ii) Search for the Clifford $C$ that takes $\ket{U}$ to $\ket{V}$.  Existence of such a Clifford is not guaranteed by virtue of $\mathcal{R}(\ket{U})=\mathcal{R}(\ket{V})$, but we found a Clifford in every instance investigated. Note also that our $T$ count is over the CNOT$+T$ basis, which is less general than the Clifford$+T$ basis but existing techniques for the latter~\cite{gosset_algorithm_2014} are impractical for more than three qubits. % This is necessary because even using special algebraic forms~\cite{gosset_algorithm_2014}, the size of the look-up tables required prohibit using Clifford$+T$ synthesis for more than three qubits whereas the CNOT$+T$ technique~\cite{amy_t-count_2016} is practical for higher numbers of qubits. 
The two methods (Clifford$+T$ and CNOT$+T$) give the same $T$ count for $CCZ$ and it is an interesting open question whether they always agree on the $T$ cost of synthesizing a third-level gate.  We list, in compact notation, a few of the new synthesis results and the $T$ savings (more are provided in Supplementary Material). For example the $CCZ$ construction discussed above would be represented as
\begin{align}
CCZ_{123}  &\overset{7 \rightarrow 4}{\longrightarrow} CS_{12}CCZ_{123} ,
\end{align}
where the subscripts denote the qubits on which a third-level gate acts  and the numbers above the arrow denote the $T$ cost. Other examples include
\begin{align}
CCZ_{123}  &\overset{7 \rightarrow 4}{\longrightarrow} CS_{12}CS_{13},\\
CCZ_{123,145}  &\overset{11 \rightarrow 8}{\longrightarrow} CS_{12,13,14,15},\\
T_{1,2,3}CS_{12,23,13}  &\overset{6 \rightarrow 5}{\longrightarrow} T_{2,3}CS_{12,23,13}.
%%T_{1,2,3}CS_{12,23,13}  &\overset{6 \rightarrow 5}{\longrightarrow} T_{1}CS_{12,13}\\
%T_{1,2}CCZ_{345}  &\overset{8 \rightarrow 6}{\longrightarrow}T_{1,2}CS_{35,45} \\
%T_{1,2,5}CCZ_{345}  &\overset{8 \rightarrow 7}{\longrightarrow}T_{1,2,5}CS_{35,45} \\
%T_{1,2,3,4}CS_{23,24,34}  &\overset{7 \rightarrow 6}{\longrightarrow} T_{1,4}CS_{24,34} \\
%CS_{12}CCZ_{345} &\overset{9 \rightarrow 7}{\longrightarrow}CS_{12,35,45}\\
%CS_{12,25}CCZ_{345} &\overset{9 \rightarrow 8}{\longrightarrow}CS_{14,25,35,45}\\
%T_{4,5}CS_{12,25}CCZ_{345} &\overset{9 \rightarrow 8}{\longrightarrow}T_4 CS_{12,25}CCZ_{345}\\
%T_{5}CS_{12,25}CCZ_{345} &\overset{9 \rightarrow 7}{\longrightarrow}T_{5}CS_{14,25,35,45}
\end{align}

\noindent\textit{Discussion.-} By reformulating robustness as an optimization in \eqref{eqn:RobustLP} this facilitates a comparison with recent related works see Table~\ref{Tab:Other_work}. For qudit-based computation, Veitch et al.~\cite{veitch_resource_2014} showed that sum-negativity $sn(\rho)$ of a state's discrete Wigner function was a well-defined resource and Pashayan et al.~\cite{pashayan_estimating_2015} showed how the run time of a Monte-Carlo type sampling algorithm was slower by a factor quadratic in the size of the sum-negativity. In the qudit setting the natural choice for the columns of $A$ in Eq.~\eqref{eqn:RobustLP} are the vertices of the Wigner polytope (a larger, but more geometrically simple object than the stabilizer polytope), and phase point operators form a natural operator basis. With these choices, $b$ is a vectorised version of the Wigner representation of $\rho$ and the matrix $A$ becomes the identity matrix. $\mathcal{R}(\rho)$ is simply the sum-negativity (equal to the $\ell_1$ norm) of the Wigner quasiprobability distribution associated with $\rho$ . In other words, sum-negativity is just robustness relative to the set of operators with non-negative discrete Wigner function.  Unlike our approach, the discrete Wigner-function is not easily adapted to qubits (although see \cite{delfosse_wigner_2015,raussendorf_contextuality_2015}).

In work by Bravyi, Smith and Smolin~\cite{bravyi_trading_2016} $t$-fold copies of $\ket{H}$ are decomposed as linear combinations of stabilizer vectors~\cite{garcia_geometry_2014}; the number of terms $\chi$---the stabilizer Schmidt rank---in the decomposition quantifies the simulation overhead. Finding the optimal decomposition is an $\ell_0$ minimization ($||x||_0=|\{i: x_i \neq 0\}|$), which is non-convex and NP-hard, limiting calculations to small number of qubits. Bravyi and Gossett~\cite{bravyi_improved_2016} extended this analysis by efficiently finding approximate decompositions that are still sufficient for the task of simulating the outcome of a quantum algorithm. This approximation precludes the possibility of ordering states by the amount of resource, however.  We note that it is well known in the signal processing literature \cite{candes_decoding_2005} that the solution to $\ell_1$ minimization also provides a (qualitatively) good solution for $\ell_0$ minimization.

\begin{table}[h]
\centering
\begin{tabular}{rccc}
\hline
Resource & $\ket{\psi}\in \mathbb{C}^{d}$ & $\rho\in B(\mathcal{H})$\\
\hline
$\chi(\ket{\psi})=||x||_0$  & BSS\cite{bravyi_trading_2016}, BG\cite{bravyi_improved_2016} \\
$sn(\rho)=||x||_1$   & & PWB\cite{pashayan_estimating_2015}, VMGE\cite{veitch_resource_2014} \\
$\mathcal{R}(\rho)=||x||_1$  & & This Work\\
\hline
\end{tabular}
\caption{\label{Tab:Other_work} Restatement of related work in terms of norm-minimizing solutions of a system of equations $Ax=b$. The amount of resource in an ancillary state $\ket{\psi}$ or $\rho$ quantifies the classical simulation overhead. In the first and third lines the columns of $A$ are $n$-qubit stabilizer states (as complex vectors or generalized Bloch vectors, respectively). In the second line the columns of $A$ are extreme points of the Wigner polytope.}
\end{table}

In Bravyi, Smith and Smolin~\cite{bravyi_trading_2016} it is conjectured that $\ket{H^{\otimes t}}$ has the smallest $\chi$ of all $t$-fold copies of a single-qubit non-stabilizer state. In our work we see that $\ket{H^{\otimes t}}$ have relatively large robustness. This is curious and worthy of further investigation but is also strongly reminiscent of~\cite{van_den_nest_universal_2013} where the (entanglement) Schmidt rank is seen to disagree with almost every other continuous entanglement measure. A related open problem is to reconcile the fact that small angle ancillae $(1,e^{i\phi\approx 0})/\sqrt{2}$ are cheap in our framework, yet are harder to synthesize over the Clifford$+T$ gate~\cite{ross_optimal_2014} set and harder to fault-tolerantly distill~\cite{duclos-cianci_distillation_2013,campbell_efficient_2016}. Considerations such as this suggest that a combination of both the stabilizer Schmidt rank and robustness pictures of magic could prove useful.

\noindent\textit{Acknowledgments}
The authors thank Naomi Nickerson for reviving our
interest in this problem and acknowledge funding from
Engineering and Physical Sciences Research Council
(Grant No. EP/M024261/1).

\bibliography{Robustness_main}

%merlin.mbs apsrev4-1.bst 2010-07-25 4.21a (PWD, AO, DPC) hacked
%Control: key (0)
%Control: author (8) initials jnrlst
%Control: editor formatted (1) identically to author
%Control: production of article title (-1) disabled
%Control: page (0) single
%Control: year (1) truncated
%Control: production of eprint (0) enabled
\begin{thebibliography}{52}%
\makeatletter
\providecommand \@ifxundefined [1]{%
 \@ifx{#1\undefined}
}%
\providecommand \@ifnum [1]{%
 \ifnum #1\expandafter \@firstoftwo
 \else \expandafter \@secondoftwo
 \fi
}%
\providecommand \@ifx [1]{%
 \ifx #1\expandafter \@firstoftwo
 \else \expandafter \@secondoftwo
 \fi
}%
\providecommand \natexlab [1]{#1}%
\providecommand \enquote  [1]{``#1''}%
\providecommand \bibnamefont  [1]{#1}%
\providecommand \bibfnamefont [1]{#1}%
\providecommand \citenamefont [1]{#1}%
\providecommand \href@noop [0]{\@secondoftwo}%
\providecommand \href [0]{\begingroup \@sanitize@url \@href}%
\providecommand \@href[1]{\@@startlink{#1}\@@href}%
\providecommand \@@href[1]{\endgroup#1\@@endlink}%
\providecommand \@sanitize@url [0]{\catcode `\\12\catcode `\$12\catcode
  `\&12\catcode `\#12\catcode `\^12\catcode `\_12\catcode `\%12\relax}%
\providecommand \@@startlink[1]{}%
\providecommand \@@endlink[0]{}%
\providecommand \url  [0]{\begingroup\@sanitize@url \@url }%
\providecommand \@url [1]{\endgroup\@href {#1}{\urlprefix }}%
\providecommand \urlprefix  [0]{URL }%
\providecommand \Eprint [0]{\href }%
\providecommand \doibase [0]{http://dx.doi.org/}%
\providecommand \selectlanguage [0]{\@gobble}%
\providecommand \bibinfo  [0]{\@secondoftwo}%
\providecommand \bibfield  [0]{\@secondoftwo}%
\providecommand \translation [1]{[#1]}%
\providecommand \BibitemOpen [0]{}%
\providecommand \bibitemStop [0]{}%
\providecommand \bibitemNoStop [0]{.\EOS\space}%
\providecommand \EOS [0]{\spacefactor3000\relax}%
\providecommand \BibitemShut  [1]{\csname bibitem#1\endcsname}%
\let\auto@bib@innerbib\@empty
%</preamble>
\bibitem [{\citenamefont {Brand{\~a}o}\ and\ \citenamefont
  {Gour}(2015)}]{brandao_reversible_2015}%
  \BibitemOpen
  \bibfield  {author} {\bibinfo {author} {\bibfnamefont {F.}~\bibnamefont
  {Brand{\~a}o}}\ and\ \bibinfo {author} {\bibfnamefont {G.}~\bibnamefont
  {Gour}},\ }\href {\doibase 10.1103/PhysRevLett.115.070503} {\bibfield
  {journal} {\bibinfo  {journal} {Phys. Rev. Lett.}\ }\textbf {\bibinfo
  {volume} {115}},\ \bibinfo {pages} {070503} (\bibinfo {year}
  {2015})}\BibitemShut {NoStop}%
\bibitem [{\citenamefont {Coecke}\ \emph {et~al.}(2016)\citenamefont {Coecke},
  \citenamefont {Fritz},\ and\ \citenamefont
  {Spekkens}}]{coecke_mathematical_2016}%
  \BibitemOpen
  \bibfield  {author} {\bibinfo {author} {\bibfnamefont {B.}~\bibnamefont
  {Coecke}}, \bibinfo {author} {\bibfnamefont {T.}~\bibnamefont {Fritz}}, \
  and\ \bibinfo {author} {\bibfnamefont {R.~W.}\ \bibnamefont {Spekkens}},\
  }\href {\doibase 10.1016/j.ic.2016.02.008} {\bibfield  {journal} {\bibinfo
  {journal} {Information and Computation}\ } (\bibinfo {year} {2016}),\
  10.1016/j.ic.2016.02.008}\BibitemShut {NoStop}%
\bibitem [{\citenamefont {Grudka}\ \emph {et~al.}(2014)\citenamefont {Grudka},
  \citenamefont {Horodecki}, \citenamefont {Horodecki}, \citenamefont
  {Horodecki}, \citenamefont {Horodecki}, \citenamefont {Joshi}, \citenamefont
  {K{\l}obus},\ and\ \citenamefont {W{\'o}jcik}}]{grudka_quantifying_2014}%
  \BibitemOpen
  \bibfield  {author} {\bibinfo {author} {\bibfnamefont {A.}~\bibnamefont
  {Grudka}}, \bibinfo {author} {\bibfnamefont {K.}~\bibnamefont {Horodecki}},
  \bibinfo {author} {\bibfnamefont {M.}~\bibnamefont {Horodecki}}, \bibinfo
  {author} {\bibfnamefont {P.}~\bibnamefont {Horodecki}}, \bibinfo {author}
  {\bibfnamefont {R.}~\bibnamefont {Horodecki}}, \bibinfo {author}
  {\bibfnamefont {P.}~\bibnamefont {Joshi}}, \bibinfo {author} {\bibfnamefont
  {W.}~\bibnamefont {K{\l}obus}}, \ and\ \bibinfo {author} {\bibfnamefont
  {A.}~\bibnamefont {W{\'o}jcik}},\ }\href {\doibase
  10.1103/PhysRevLett.112.120401} {\bibfield  {journal} {\bibinfo  {journal}
  {Phys. Rev. Lett.}\ }\textbf {\bibinfo {volume} {112}},\ \bibinfo {pages}
  {120401} (\bibinfo {year} {2014})}\BibitemShut {NoStop}%
\bibitem [{\citenamefont {Horodecki}\ and\ \citenamefont
  {Oppenheim}(2013)}]{Horodecki:2013}%
  \BibitemOpen
  \bibfield  {author} {\bibinfo {author} {\bibfnamefont {M.}~\bibnamefont
  {Horodecki}}\ and\ \bibinfo {author} {\bibfnamefont {J.}~\bibnamefont
  {Oppenheim}},\ }\href {\doibase 10.1142/S0217979213450197} {\bibfield
  {journal} {\bibinfo  {journal} {Int. J. Mod. Phys. B}\ }\textbf {\bibinfo
  {volume} {27}},\ \bibinfo {pages} {1345019} (\bibinfo {year}
  {2013})}\BibitemShut {NoStop}%
\bibitem [{\citenamefont {Napoli}\ \emph {et~al.}(2016)\citenamefont {Napoli},
  \citenamefont {Bromley}, \citenamefont {Cianciaruso}, \citenamefont {Piani},
  \citenamefont {Johnston},\ and\ \citenamefont
  {Adesso}}]{napoli_robustness_2016}%
  \BibitemOpen
  \bibfield  {author} {\bibinfo {author} {\bibfnamefont {C.}~\bibnamefont
  {Napoli}}, \bibinfo {author} {\bibfnamefont {T.~R.}\ \bibnamefont {Bromley}},
  \bibinfo {author} {\bibfnamefont {M.}~\bibnamefont {Cianciaruso}}, \bibinfo
  {author} {\bibfnamefont {M.}~\bibnamefont {Piani}}, \bibinfo {author}
  {\bibfnamefont {N.}~\bibnamefont {Johnston}}, \ and\ \bibinfo {author}
  {\bibfnamefont {G.}~\bibnamefont {Adesso}},\ }\href {\doibase
  10.1103/PhysRevLett.116.150502} {\bibfield  {journal} {\bibinfo  {journal}
  {Phys. Rev. Lett.}\ }\textbf {\bibinfo {volume} {116}},\ \bibinfo {pages}
  {150502} (\bibinfo {year} {2016})}\BibitemShut {NoStop}%
\bibitem [{\citenamefont {Stahlke}(2014)}]{stahlke_quantum_2014}%
  \BibitemOpen
  \bibfield  {author} {\bibinfo {author} {\bibfnamefont {D.}~\bibnamefont
  {Stahlke}},\ }\href {\doibase 10.1103/PhysRevA.90.022302} {\bibfield
  {journal} {\bibinfo  {journal} {Phys. Rev. A}\ }\textbf {\bibinfo {volume}
  {90}},\ \bibinfo {pages} {022302} (\bibinfo {year} {2014})}\BibitemShut
  {NoStop}%
\bibitem [{\citenamefont {Veitch}\ \emph {et~al.}(2014)\citenamefont {Veitch},
  \citenamefont {Mousavian}, \citenamefont {Gottesman},\ and\ \citenamefont
  {Emerson}}]{veitch_resource_2014}%
  \BibitemOpen
  \bibfield  {author} {\bibinfo {author} {\bibfnamefont {V.}~\bibnamefont
  {Veitch}}, \bibinfo {author} {\bibfnamefont {S.~A.~H.}\ \bibnamefont
  {Mousavian}}, \bibinfo {author} {\bibfnamefont {D.}~\bibnamefont
  {Gottesman}}, \ and\ \bibinfo {author} {\bibfnamefont {J.}~\bibnamefont
  {Emerson}},\ }\href {\doibase 10.1088/1367-2630/16/1/013009} {\bibfield
  {journal} {\bibinfo  {journal} {New J. Phys.}\ }\textbf {\bibinfo {volume}
  {16}},\ \bibinfo {pages} {013009} (\bibinfo {year} {2014})}\BibitemShut
  {NoStop}%
\bibitem [{\citenamefont {Vidal}\ and\ \citenamefont
  {Tarrach}(1999)}]{vidal_robustness_1999}%
  \BibitemOpen
  \bibfield  {author} {\bibinfo {author} {\bibfnamefont {G.}~\bibnamefont
  {Vidal}}\ and\ \bibinfo {author} {\bibfnamefont {R.}~\bibnamefont
  {Tarrach}},\ }\href {\doibase 10.1103/PhysRevA.59.141} {\bibfield  {journal}
  {\bibinfo  {journal} {Phys. Rev. A}\ }\textbf {\bibinfo {volume} {59}},\
  \bibinfo {pages} {141} (\bibinfo {year} {1999})}\BibitemShut {NoStop}%
\bibitem [{\citenamefont {Bravyi}\ and\ \citenamefont
  {Kitaev}(2005)}]{bravyi_universal_2005}%
  \BibitemOpen
  \bibfield  {author} {\bibinfo {author} {\bibfnamefont {S.}~\bibnamefont
  {Bravyi}}\ and\ \bibinfo {author} {\bibfnamefont {A.}~\bibnamefont
  {Kitaev}},\ }\href {\doibase 10.1103/PhysRevA.71.022316} {\bibfield
  {journal} {\bibinfo  {journal} {Phys. Rev. A}\ }\textbf {\bibinfo {volume}
  {71}},\ \bibinfo {pages} {022316} (\bibinfo {year} {2005})}\BibitemShut
  {NoStop}%
\bibitem [{\citenamefont {Knill}(2005)}]{knill_quantum_2005}%
  \BibitemOpen
  \bibfield  {author} {\bibinfo {author} {\bibfnamefont {E.}~\bibnamefont
  {Knill}},\ }\href {\doibase 10.1038/nature03350} {\bibfield  {journal}
  {\bibinfo  {journal} {Nature}\ }\textbf {\bibinfo {volume} {434}},\ \bibinfo
  {pages} {39} (\bibinfo {year} {2005})}\BibitemShut {NoStop}%
\bibitem [{\citenamefont {Campbell}\ \emph {et~al.}(2016)\citenamefont
  {Campbell}, \citenamefont {Terhal},\ and\ \citenamefont
  {Vuillot}}]{campbell_steep_2016}%
  \BibitemOpen
  \bibfield  {author} {\bibinfo {author} {\bibfnamefont {E.~T.}\ \bibnamefont
  {Campbell}}, \bibinfo {author} {\bibfnamefont {B.~M.}\ \bibnamefont
  {Terhal}}, \ and\ \bibinfo {author} {\bibfnamefont {C.}~\bibnamefont
  {Vuillot}},\ }\href@noop {} {\bibfield  {journal} {\bibinfo  {journal}
  {arXiv:1612.07330 [quant-ph]}\ } (\bibinfo {year} {2016})},\ \Eprint
  {http://arxiv.org/abs/1612.07330} {arXiv:1612.07330 [quant-ph]} \BibitemShut
  {NoStop}%
\bibitem [{\citenamefont {Howard}\ \emph {et~al.}(2014)\citenamefont {Howard},
  \citenamefont {Wallman}, \citenamefont {Veitch},\ and\ \citenamefont
  {Emerson}}]{howard_contextuality_2014}%
  \BibitemOpen
  \bibfield  {author} {\bibinfo {author} {\bibfnamefont {M.}~\bibnamefont
  {Howard}}, \bibinfo {author} {\bibfnamefont {J.}~\bibnamefont {Wallman}},
  \bibinfo {author} {\bibfnamefont {V.}~\bibnamefont {Veitch}}, \ and\ \bibinfo
  {author} {\bibfnamefont {J.}~\bibnamefont {Emerson}},\ }\href {\doibase
  10.1038/nature13460} {\bibfield  {journal} {\bibinfo  {journal} {Nature}\
  }\textbf {\bibinfo {volume} {510}},\ \bibinfo {pages} {351} (\bibinfo {year}
  {2014})}\BibitemShut {NoStop}%
\bibitem [{\citenamefont {Delfosse}\ \emph {et~al.}(2016)\citenamefont
  {Delfosse}, \citenamefont {Okay}, \citenamefont {Bermejo-Vega}, \citenamefont
  {Browne},\ and\ \citenamefont {Raussendorf}}]{delfosse_equivalence_2016}%
  \BibitemOpen
  \bibfield  {author} {\bibinfo {author} {\bibfnamefont {N.}~\bibnamefont
  {Delfosse}}, \bibinfo {author} {\bibfnamefont {C.}~\bibnamefont {Okay}},
  \bibinfo {author} {\bibfnamefont {J.}~\bibnamefont {Bermejo-Vega}}, \bibinfo
  {author} {\bibfnamefont {D.~E.}\ \bibnamefont {Browne}}, \ and\ \bibinfo
  {author} {\bibfnamefont {R.}~\bibnamefont {Raussendorf}},\ }\href@noop {}
  {\bibfield  {journal} {\bibinfo  {journal} {arXiv:1610.07093 [quant-ph]}\ }
  (\bibinfo {year} {2016})},\ \Eprint {http://arxiv.org/abs/1610.07093}
  {arXiv:1610.07093 [quant-ph]} \BibitemShut {NoStop}%
\bibitem [{\citenamefont {Veitch}\ \emph {et~al.}(2012)\citenamefont {Veitch},
  \citenamefont {Ferrie}, \citenamefont {Gross},\ and\ \citenamefont
  {Emerson}}]{veitch_negative_2012}%
  \BibitemOpen
  \bibfield  {author} {\bibinfo {author} {\bibfnamefont {V.}~\bibnamefont
  {Veitch}}, \bibinfo {author} {\bibfnamefont {C.}~\bibnamefont {Ferrie}},
  \bibinfo {author} {\bibfnamefont {D.}~\bibnamefont {Gross}}, \ and\ \bibinfo
  {author} {\bibfnamefont {J.}~\bibnamefont {Emerson}},\ }\href {\doibase
  10.1088/1367-2630/14/11/113011} {\bibfield  {journal} {\bibinfo  {journal}
  {New J. Phys.}\ }\textbf {\bibinfo {volume} {14}},\ \bibinfo {pages} {113011}
  (\bibinfo {year} {2012})}\BibitemShut {NoStop}%
\bibitem [{\citenamefont {Mari}\ and\ \citenamefont
  {Eisert}(2012)}]{mari_positive_2012}%
  \BibitemOpen
  \bibfield  {author} {\bibinfo {author} {\bibfnamefont {A.}~\bibnamefont
  {Mari}}\ and\ \bibinfo {author} {\bibfnamefont {J.}~\bibnamefont {Eisert}},\
  }\href {\doibase 10.1103/PhysRevLett.109.230503} {\bibfield  {journal}
  {\bibinfo  {journal} {Phys. Rev. Lett.}\ }\textbf {\bibinfo {volume} {109}},\
  \bibinfo {pages} {230503} (\bibinfo {year} {2012})}\BibitemShut {NoStop}%
\bibitem [{\citenamefont {Gross}(2006)}]{gross_hudsons_2006}%
  \BibitemOpen
  \bibfield  {author} {\bibinfo {author} {\bibfnamefont {D.}~\bibnamefont
  {Gross}},\ }\href {\doibase 10.1063/1.2393152} {\bibfield  {journal}
  {\bibinfo  {journal} {Journal of Mathematical Physics}\ }\textbf {\bibinfo
  {volume} {47}},\ \bibinfo {pages} {122107} (\bibinfo {year}
  {2006})}\BibitemShut {NoStop}%
\bibitem [{\citenamefont {Aaronson}\ and\ \citenamefont
  {Gottesman}(2004)}]{aaronson_improved_2004}%
  \BibitemOpen
  \bibfield  {author} {\bibinfo {author} {\bibfnamefont {S.}~\bibnamefont
  {Aaronson}}\ and\ \bibinfo {author} {\bibfnamefont {D.}~\bibnamefont
  {Gottesman}},\ }\href {\doibase 10.1103/PhysRevA.70.052328} {\bibfield
  {journal} {\bibinfo  {journal} {Phys. Rev. A}\ }\textbf {\bibinfo {volume}
  {70}},\ \bibinfo {pages} {052328} (\bibinfo {year} {2004})}\BibitemShut
  {NoStop}%
\bibitem [{\citenamefont {Bravyi}\ and\ \citenamefont
  {Gosset}(2016)}]{bravyi_improved_2016}%
  \BibitemOpen
  \bibfield  {author} {\bibinfo {author} {\bibfnamefont {S.}~\bibnamefont
  {Bravyi}}\ and\ \bibinfo {author} {\bibfnamefont {D.}~\bibnamefont
  {Gosset}},\ }\href {\doibase 10.1103/PhysRevLett.116.250501} {\bibfield
  {journal} {\bibinfo  {journal} {Phys. Rev. Lett.}\ }\textbf {\bibinfo
  {volume} {116}},\ \bibinfo {pages} {250501} (\bibinfo {year}
  {2016})}\BibitemShut {NoStop}%
\bibitem [{\citenamefont {Bravyi}\ \emph {et~al.}(2016)\citenamefont {Bravyi},
  \citenamefont {Smith},\ and\ \citenamefont {Smolin}}]{bravyi_trading_2016}%
  \BibitemOpen
  \bibfield  {author} {\bibinfo {author} {\bibfnamefont {S.}~\bibnamefont
  {Bravyi}}, \bibinfo {author} {\bibfnamefont {G.}~\bibnamefont {Smith}}, \
  and\ \bibinfo {author} {\bibfnamefont {J.~A.}\ \bibnamefont {Smolin}},\
  }\href {\doibase 10.1103/PhysRevX.6.021043} {\bibfield  {journal} {\bibinfo
  {journal} {Phys. Rev. X}\ }\textbf {\bibinfo {volume} {6}},\ \bibinfo {pages}
  {021043} (\bibinfo {year} {2016})}\BibitemShut {NoStop}%
\bibitem [{\citenamefont {Garc{\'\i}a}\ \emph {et~al.}(2014)\citenamefont
  {Garc{\'\i}a}, \citenamefont {Markov},\ and\ \citenamefont
  {Cross}}]{garcia_geometry_2014}%
  \BibitemOpen
  \bibfield  {author} {\bibinfo {author} {\bibfnamefont {H.~J.}\ \bibnamefont
  {Garc{\'\i}a}}, \bibinfo {author} {\bibfnamefont {I.~L.}\ \bibnamefont
  {Markov}}, \ and\ \bibinfo {author} {\bibfnamefont {A.~W.}\ \bibnamefont
  {Cross}},\ }\href@noop {} {\bibfield  {journal} {\bibinfo  {journal} {Quantum
  Information \& Computation}\ }\textbf {\bibinfo {volume} {14}},\ \bibinfo
  {pages} {683} (\bibinfo {year} {2014})}\BibitemShut {NoStop}%
\bibitem [{\citenamefont {Pashayan}\ \emph {et~al.}(2015)\citenamefont
  {Pashayan}, \citenamefont {Wallman},\ and\ \citenamefont
  {Bartlett}}]{pashayan_estimating_2015}%
  \BibitemOpen
  \bibfield  {author} {\bibinfo {author} {\bibfnamefont {H.}~\bibnamefont
  {Pashayan}}, \bibinfo {author} {\bibfnamefont {J.~J.}\ \bibnamefont
  {Wallman}}, \ and\ \bibinfo {author} {\bibfnamefont {S.~D.}\ \bibnamefont
  {Bartlett}},\ }\href {\doibase 10.1103/PhysRevLett.115.070501} {\bibfield
  {journal} {\bibinfo  {journal} {Phys. Rev. Lett.}\ }\textbf {\bibinfo
  {volume} {115}},\ \bibinfo {pages} {070501} (\bibinfo {year}
  {2015})}\BibitemShut {NoStop}%
\bibitem [{\citenamefont {Amy}\ \emph {et~al.}(2014)\citenamefont {Amy},
  \citenamefont {Maslov},\ and\ \citenamefont
  {Mosca}}]{amy_polynomial-time_2014}%
  \BibitemOpen
  \bibfield  {author} {\bibinfo {author} {\bibfnamefont {M.}~\bibnamefont
  {Amy}}, \bibinfo {author} {\bibfnamefont {D.}~\bibnamefont {Maslov}}, \ and\
  \bibinfo {author} {\bibfnamefont {M.}~\bibnamefont {Mosca}},\ }\href
  {\doibase 10.1109/TCAD.2014.2341953} {\bibfield  {journal} {\bibinfo
  {journal} {IEEE Transactions on Computer-Aided Design of Integrated Circuits
  and Systems}\ }\textbf {\bibinfo {volume} {33}},\ \bibinfo {pages} {1476}
  (\bibinfo {year} {2014})}\BibitemShut {NoStop}%
\bibitem [{\citenamefont {Amy}\ and\ \citenamefont
  {Mosca}(2016)}]{amy_t-count_2016}%
  \BibitemOpen
  \bibfield  {author} {\bibinfo {author} {\bibfnamefont {M.}~\bibnamefont
  {Amy}}\ and\ \bibinfo {author} {\bibfnamefont {M.}~\bibnamefont {Mosca}},\
  }\href@noop {} {\bibfield  {journal} {\bibinfo  {journal} {arXiv:1601.07363
  [quant-ph]}\ } (\bibinfo {year} {2016})},\ \Eprint
  {http://arxiv.org/abs/1601.07363} {arXiv:1601.07363 [quant-ph]} \BibitemShut
  {NoStop}%
\bibitem [{\citenamefont {Gosset}\ \emph {et~al.}(2014)\citenamefont {Gosset},
  \citenamefont {Kliuchnikov}, \citenamefont {Mosca},\ and\ \citenamefont
  {Russo}}]{gosset_algorithm_2014}%
  \BibitemOpen
  \bibfield  {author} {\bibinfo {author} {\bibfnamefont {D.}~\bibnamefont
  {Gosset}}, \bibinfo {author} {\bibfnamefont {V.}~\bibnamefont {Kliuchnikov}},
  \bibinfo {author} {\bibfnamefont {M.}~\bibnamefont {Mosca}}, \ and\ \bibinfo
  {author} {\bibfnamefont {V.}~\bibnamefont {Russo}},\ }\href@noop {}
  {\bibfield  {journal} {\bibinfo  {journal} {Quantum Information \&
  Computation}\ }\textbf {\bibinfo {volume} {14}},\ \bibinfo {pages} {1261}
  (\bibinfo {year} {2014})}\BibitemShut {NoStop}%
\bibitem [{\citenamefont {Bocharov}\ \emph {et~al.}(2013)\citenamefont
  {Bocharov}, \citenamefont {Gurevich},\ and\ \citenamefont
  {Svore}}]{bocharov_efficient_2013}%
  \BibitemOpen
  \bibfield  {author} {\bibinfo {author} {\bibfnamefont {A.}~\bibnamefont
  {Bocharov}}, \bibinfo {author} {\bibfnamefont {Y.}~\bibnamefont {Gurevich}},
  \ and\ \bibinfo {author} {\bibfnamefont {K.~M.}\ \bibnamefont {Svore}},\
  }\href {\doibase 10.1103/PhysRevA.88.012313} {\bibfield  {journal} {\bibinfo
  {journal} {Phys. Rev. A}\ }\textbf {\bibinfo {volume} {88}},\ \bibinfo
  {pages} {012313} (\bibinfo {year} {2013})}\BibitemShut {NoStop}%
\bibitem [{\citenamefont {Duclos-Cianci}\ and\ \citenamefont
  {Svore}(2013)}]{duclos-cianci_distillation_2013}%
  \BibitemOpen
  \bibfield  {author} {\bibinfo {author} {\bibfnamefont {G.}~\bibnamefont
  {Duclos-Cianci}}\ and\ \bibinfo {author} {\bibfnamefont {K.~M.}\ \bibnamefont
  {Svore}},\ }\href {\doibase 10.1103/PhysRevA.88.042325} {\bibfield  {journal}
  {\bibinfo  {journal} {Phys. Rev. A}\ }\textbf {\bibinfo {volume} {88}},\
  \bibinfo {pages} {042325} (\bibinfo {year} {2013})}\BibitemShut {NoStop}%
\bibitem [{\citenamefont {Paetznick}\ and\ \citenamefont
  {Svore}(2014)}]{paetznick_repeat-until-success:_2014}%
  \BibitemOpen
  \bibfield  {author} {\bibinfo {author} {\bibfnamefont {A.}~\bibnamefont
  {Paetznick}}\ and\ \bibinfo {author} {\bibfnamefont {K.~M.}\ \bibnamefont
  {Svore}},\ }\href@noop {} {\bibfield  {journal} {\bibinfo  {journal} {Quantum
  Information \& Computation}\ }\textbf {\bibinfo {volume} {14}},\ \bibinfo
  {pages} {1277} (\bibinfo {year} {2014})}\BibitemShut {NoStop}%
\bibitem [{\citenamefont {Ross}\ and\ \citenamefont
  {Selinger}(2014)}]{ross_optimal_2014}%
  \BibitemOpen
  \bibfield  {author} {\bibinfo {author} {\bibfnamefont {N.~J.}\ \bibnamefont
  {Ross}}\ and\ \bibinfo {author} {\bibfnamefont {P.}~\bibnamefont
  {Selinger}},\ }\href@noop {} {\bibfield  {journal} {\bibinfo  {journal}
  {arXiv:1403.2975 [quant-ph]}\ } (\bibinfo {year} {2014})},\ \Eprint
  {http://arxiv.org/abs/1403.2975} {arXiv:1403.2975 [quant-ph]} \BibitemShut
  {NoStop}%
\bibitem [{\citenamefont {Selinger}(2013)}]{selinger_quantum_2013}%
  \BibitemOpen
  \bibfield  {author} {\bibinfo {author} {\bibfnamefont {P.}~\bibnamefont
  {Selinger}},\ }\href {\doibase 10.1103/PhysRevA.87.042302} {\bibfield
  {journal} {\bibinfo  {journal} {Phys. Rev. A}\ }\textbf {\bibinfo {volume}
  {87}},\ \bibinfo {pages} {042302} (\bibinfo {year} {2013})}\BibitemShut
  {NoStop}%
\bibitem [{\citenamefont {Wiebe}\ and\ \citenamefont
  {Roetteler}(2016)}]{wiebe_quantum_2016}%
  \BibitemOpen
  \bibfield  {author} {\bibinfo {author} {\bibfnamefont {N.}~\bibnamefont
  {Wiebe}}\ and\ \bibinfo {author} {\bibfnamefont {M.}~\bibnamefont
  {Roetteler}},\ }\href@noop {} {\bibfield  {journal} {\bibinfo  {journal}
  {Quantum Information and Communication}\ }\textbf {\bibinfo {volume} {16}},\
  \bibinfo {pages} {134} (\bibinfo {year} {2016})}\BibitemShut {NoStop}%
\bibitem [{\citenamefont {Jones}(2013)}]{jones_low-overhead_2013}%
  \BibitemOpen
  \bibfield  {author} {\bibinfo {author} {\bibfnamefont {C.}~\bibnamefont
  {Jones}},\ }\href {\doibase 10.1103/PhysRevA.87.022328} {\bibfield  {journal}
  {\bibinfo  {journal} {Phys. Rev. A}\ }\textbf {\bibinfo {volume} {87}},\
  \bibinfo {pages} {022328} (\bibinfo {year} {2013})}\BibitemShut {NoStop}%
\bibitem [{\citenamefont {Andersson}\ \emph {et~al.}(2015)\citenamefont
  {Andersson}, \citenamefont {Bengtsson}, \citenamefont {Blanchfield},\ and\
  \citenamefont {Dang}}]{andersson_states_2015}%
  \BibitemOpen
  \bibfield  {author} {\bibinfo {author} {\bibfnamefont {D.}~\bibnamefont
  {Andersson}}, \bibinfo {author} {\bibfnamefont {I.}~\bibnamefont
  {Bengtsson}}, \bibinfo {author} {\bibfnamefont {K.}~\bibnamefont
  {Blanchfield}}, \ and\ \bibinfo {author} {\bibfnamefont {H.~B.}\ \bibnamefont
  {Dang}},\ }\href {\doibase 10.1088/1751-8113/48/34/345301} {\bibfield
  {journal} {\bibinfo  {journal} {J. Phys. A}\ }\textbf {\bibinfo {volume}
  {48}},\ \bibinfo {pages} {345301} (\bibinfo {year} {2015})}\BibitemShut
  {NoStop}%
\bibitem [{\citenamefont {Reichardt}(2009)}]{reichardt_quantum_2009}%
  \BibitemOpen
  \bibfield  {author} {\bibinfo {author} {\bibfnamefont {B.~W.}\ \bibnamefont
  {Reichardt}},\ }\href@noop {} {\bibfield  {journal} {\bibinfo  {journal}
  {Quantum Information \& Computation}\ }\textbf {\bibinfo {volume} {9}},\
  \bibinfo {pages} {1030} (\bibinfo {year} {2009})}\BibitemShut {NoStop}%
\bibitem [{\citenamefont {Campbell}(2011)}]{campbell_catalysis_2011}%
  \BibitemOpen
  \bibfield  {author} {\bibinfo {author} {\bibfnamefont {E.~T.}\ \bibnamefont
  {Campbell}},\ }\href {\doibase 10.1103/PhysRevA.83.032317} {\bibfield
  {journal} {\bibinfo  {journal} {Phys. Rev. A}\ }\textbf {\bibinfo {volume}
  {83}},\ \bibinfo {pages} {032317} (\bibinfo {year} {2011})}\BibitemShut
  {NoStop}%
\bibitem [{\citenamefont {Harrow}\ and\ \citenamefont
  {Nielsen}(2003)}]{harrow_robustness_2003}%
  \BibitemOpen
  \bibfield  {author} {\bibinfo {author} {\bibfnamefont {A.~W.}\ \bibnamefont
  {Harrow}}\ and\ \bibinfo {author} {\bibfnamefont {M.~A.}\ \bibnamefont
  {Nielsen}},\ }\href {\doibase 10.1103/PhysRevA.68.012308} {\bibfield
  {journal} {\bibinfo  {journal} {Phys. Rev. A}\ }\textbf {\bibinfo {volume}
  {68}},\ \bibinfo {pages} {012308} (\bibinfo {year} {2003})}\BibitemShut
  {NoStop}%
\bibitem [{\citenamefont {{van Dam}}\ and\ \citenamefont
  {Howard}(2011)}]{van_dam_noise_2011}%
  \BibitemOpen
  \bibfield  {author} {\bibinfo {author} {\bibfnamefont {W.}~\bibnamefont {{van
  Dam}}}\ and\ \bibinfo {author} {\bibfnamefont {M.}~\bibnamefont {Howard}},\
  }\href {\doibase 10.1103/PhysRevA.83.032310} {\bibfield  {journal} {\bibinfo
  {journal} {Phys. Rev. A}\ }\textbf {\bibinfo {volume} {83}},\ \bibinfo
  {pages} {032310} (\bibinfo {year} {2011})}\BibitemShut {NoStop}%
\bibitem [{\citenamefont {Hoggar}(1998)}]{hoggar_64_1998}%
  \BibitemOpen
  \bibfield  {author} {\bibinfo {author} {\bibfnamefont {S.~G.}\ \bibnamefont
  {Hoggar}},\ }\href {\doibase 10.1023/A:1005009727232} {\bibfield  {journal}
  {\bibinfo  {journal} {Geometriae Dedicata}\ }\textbf {\bibinfo {volume}
  {69}},\ \bibinfo {pages} {287} (\bibinfo {year} {1998})}\BibitemShut
  {NoStop}%
\bibitem [{\citenamefont {Grant}\ and\ \citenamefont
  {Boyd}(2008)}]{grant_graph_2008}%
  \BibitemOpen
  \bibfield  {author} {\bibinfo {author} {\bibfnamefont {M.}~\bibnamefont
  {Grant}}\ and\ \bibinfo {author} {\bibfnamefont {S.}~\bibnamefont {Boyd}},\
  }in\ \href@noop {} {\emph {\bibinfo {booktitle} {Recent {{Advances}} in
  {{Learning}} and {{Control}}}}},\ \bibinfo {series and number} {Lecture Notes
  in Control and Information Sciences},\ \bibinfo {editor} {edited by\ \bibinfo
  {editor} {\bibfnamefont {V.}~\bibnamefont {Blondel}}, \bibinfo {editor}
  {\bibfnamefont {S.}~\bibnamefont {Boyd}}, \ and\ \bibinfo {editor}
  {\bibfnamefont {H.}~\bibnamefont {Kimura}}}\ (\bibinfo  {publisher}
  {{Springer-Verlag Limited}},\ \bibinfo {year} {2008})\ pp.\ \bibinfo {pages}
  {95--110}\BibitemShut {NoStop}%
\bibitem [{\citenamefont {Grant}\ and\ \citenamefont
  {Boyd}(2014)}]{grant_cvx:_2014}%
  \BibitemOpen
  \bibfield  {author} {\bibinfo {author} {\bibfnamefont {M.}~\bibnamefont
  {Grant}}\ and\ \bibinfo {author} {\bibfnamefont {S.}~\bibnamefont {Boyd}},\
  }\href@noop {} {\emph {\bibinfo {title} {{{CVX}}: {{Matlab Software}} for
  {{Disciplined Convex Programming}}, Version 2.1}}}\ (\bibinfo {year}
  {2014})\BibitemShut {NoStop}%
\bibitem [{Note1()}]{Note1}%
  \BibitemOpen
  \bibinfo {note} {This certificate of optimality can be used to obtain a magic
  witness---an operator whose expectation value with respect to stabilizer
  states is in the interval $[-1,1]$ and whose expectation with respect to
  $\rho $ is $\protect \mathcal {R}(\rho )$. These witnesses can be used to
  derive exact expressions for robustness as in Supplementary
  Material}\BibitemShut {NoStop}%
\bibitem [{\citenamefont {Buhrman}\ \emph {et~al.}(2006)\citenamefont
  {Buhrman}, \citenamefont {Cleve}, \citenamefont {Laurent}, \citenamefont
  {Linden}, \citenamefont {Schrijver},\ and\ \citenamefont
  {Unger}}]{buhrman_new_2006}%
  \BibitemOpen
  \bibfield  {author} {\bibinfo {author} {\bibfnamefont {H.}~\bibnamefont
  {Buhrman}}, \bibinfo {author} {\bibfnamefont {R.}~\bibnamefont {Cleve}},
  \bibinfo {author} {\bibfnamefont {M.}~\bibnamefont {Laurent}}, \bibinfo
  {author} {\bibfnamefont {N.}~\bibnamefont {Linden}}, \bibinfo {author}
  {\bibfnamefont {A.}~\bibnamefont {Schrijver}}, \ and\ \bibinfo {author}
  {\bibfnamefont {F.}~\bibnamefont {Unger}},\ }in\ \href {\doibase
  10.1109/FOCS.2006.50} {\emph {\bibinfo {booktitle} {2006 47th {{Annual IEEE
  Symposium}} on {{Foundations}} of {{Computer Science}} ({{FOCS}}'06)}}}\
  (\bibinfo {year} {2006})\ pp.\ \bibinfo {pages} {411--419}\BibitemShut
  {NoStop}%
\bibitem [{\citenamefont {Virmani}\ \emph {et~al.}(2005)\citenamefont
  {Virmani}, \citenamefont {Huelga},\ and\ \citenamefont
  {Plenio}}]{virmani_classical_2005}%
  \BibitemOpen
  \bibfield  {author} {\bibinfo {author} {\bibfnamefont {S.}~\bibnamefont
  {Virmani}}, \bibinfo {author} {\bibfnamefont {S.~F.}\ \bibnamefont {Huelga}},
  \ and\ \bibinfo {author} {\bibfnamefont {M.~B.}\ \bibnamefont {Plenio}},\
  }\href {\doibase 10.1103/PhysRevA.71.042328} {\bibfield  {journal} {\bibinfo
  {journal} {Physical Review A}\ }\textbf {\bibinfo {volume} {71}},\ \bibinfo
  {pages} {042328} (\bibinfo {year} {2005})}\BibitemShut {NoStop}%
\bibitem [{\citenamefont {Plenio}\ and\ \citenamefont
  {Virmani}(2010)}]{plenio_upper_2010}%
  \BibitemOpen
  \bibfield  {author} {\bibinfo {author} {\bibfnamefont {M.~B.}\ \bibnamefont
  {Plenio}}\ and\ \bibinfo {author} {\bibfnamefont {S.}~\bibnamefont
  {Virmani}},\ }\href {\doibase 10.1088/1367-2630/12/3/033012} {\bibfield
  {journal} {\bibinfo  {journal} {New Journal of Physics}\ }\textbf {\bibinfo
  {volume} {12}},\ \bibinfo {pages} {033012} (\bibinfo {year}
  {2010})}\BibitemShut {NoStop}%
\bibitem [{\citenamefont {Gottesman}\ and\ \citenamefont
  {Chuang}(1999)}]{gottesman_demonstrating_1999}%
  \BibitemOpen
  \bibfield  {author} {\bibinfo {author} {\bibfnamefont {D.}~\bibnamefont
  {Gottesman}}\ and\ \bibinfo {author} {\bibfnamefont {I.~L.}\ \bibnamefont
  {Chuang}},\ }\href {\doibase 10.1038/46503} {\bibfield  {journal} {\bibinfo
  {journal} {Nature}\ }\textbf {\bibinfo {volume} {402}},\ \bibinfo {pages}
  {390} (\bibinfo {year} {1999})}\BibitemShut {NoStop}%
\bibitem [{\citenamefont {Zhou}\ \emph {et~al.}(2000)\citenamefont {Zhou},
  \citenamefont {Leung},\ and\ \citenamefont {Chuang}}]{zhou_methodology_2000}%
  \BibitemOpen
  \bibfield  {author} {\bibinfo {author} {\bibfnamefont {X.}~\bibnamefont
  {Zhou}}, \bibinfo {author} {\bibfnamefont {D.~W.}\ \bibnamefont {Leung}}, \
  and\ \bibinfo {author} {\bibfnamefont {I.~L.}\ \bibnamefont {Chuang}},\
  }\href {\doibase 10.1103/PhysRevA.62.052316} {\bibfield  {journal} {\bibinfo
  {journal} {Phys. Rev. A}\ }\textbf {\bibinfo {volume} {62}},\ \bibinfo
  {pages} {052316} (\bibinfo {year} {2000})}\BibitemShut {NoStop}%
\bibitem [{\citenamefont {Campbell}\ and\ \citenamefont
  {Howard}(2016{\natexlab{a}})}]{campbell_unified_2016}%
  \BibitemOpen
  \bibfield  {author} {\bibinfo {author} {\bibfnamefont {E.~T.}\ \bibnamefont
  {Campbell}}\ and\ \bibinfo {author} {\bibfnamefont {M.}~\bibnamefont
  {Howard}},\ }\href@noop {} {\bibfield  {journal} {\bibinfo  {journal}
  {arXiv:1606.01904 [quant-ph]}\ } (\bibinfo {year} {2016}{\natexlab{a}})},\
  \Eprint {http://arxiv.org/abs/1606.01904} {arXiv:1606.01904 [quant-ph]}
  \BibitemShut {NoStop}%
\bibitem [{\citenamefont {Campbell}\ and\ \citenamefont
  {Howard}(2016{\natexlab{b}})}]{campbell_unifying_2016}%
  \BibitemOpen
  \bibfield  {author} {\bibinfo {author} {\bibfnamefont {E.~T.}\ \bibnamefont
  {Campbell}}\ and\ \bibinfo {author} {\bibfnamefont {M.}~\bibnamefont
  {Howard}},\ }\href@noop {} {\bibfield  {journal} {\bibinfo  {journal}
  {arXiv:1606.01906 [quant-ph]}\ } (\bibinfo {year} {2016}{\natexlab{b}})},\
  \Eprint {http://arxiv.org/abs/1606.01906} {arXiv:1606.01906 [quant-ph]}
  \BibitemShut {NoStop}%
\bibitem [{\citenamefont {Delfosse}\ \emph {et~al.}(2015)\citenamefont
  {Delfosse}, \citenamefont {Allard~Guerin}, \citenamefont {Bian},\ and\
  \citenamefont {Raussendorf}}]{delfosse_wigner_2015}%
  \BibitemOpen
  \bibfield  {author} {\bibinfo {author} {\bibfnamefont {N.}~\bibnamefont
  {Delfosse}}, \bibinfo {author} {\bibfnamefont {P.}~\bibnamefont
  {Allard~Guerin}}, \bibinfo {author} {\bibfnamefont {J.}~\bibnamefont {Bian}},
  \ and\ \bibinfo {author} {\bibfnamefont {R.}~\bibnamefont {Raussendorf}},\
  }\href {\doibase 10.1103/PhysRevX.5.021003} {\bibfield  {journal} {\bibinfo
  {journal} {Phys. Rev. X}\ }\textbf {\bibinfo {volume} {5}},\ \bibinfo {pages}
  {021003} (\bibinfo {year} {2015})}\BibitemShut {NoStop}%
\bibitem [{\citenamefont {Raussendorf}\ \emph {et~al.}(2015)\citenamefont
  {Raussendorf}, \citenamefont {Browne}, \citenamefont {Delfosse},
  \citenamefont {Okay},\ and\ \citenamefont
  {Bermejo-Vega}}]{raussendorf_contextuality_2015}%
  \BibitemOpen
  \bibfield  {author} {\bibinfo {author} {\bibfnamefont {R.}~\bibnamefont
  {Raussendorf}}, \bibinfo {author} {\bibfnamefont {D.~E.}\ \bibnamefont
  {Browne}}, \bibinfo {author} {\bibfnamefont {N.}~\bibnamefont {Delfosse}},
  \bibinfo {author} {\bibfnamefont {C.}~\bibnamefont {Okay}}, \ and\ \bibinfo
  {author} {\bibfnamefont {J.}~\bibnamefont {Bermejo-Vega}},\ }\href@noop {}
  {\bibfield  {journal} {\bibinfo  {journal} {arXiv:1511.08506 [quant-ph]}\ }
  (\bibinfo {year} {2015})},\ \Eprint {http://arxiv.org/abs/1511.08506}
  {arXiv:1511.08506 [quant-ph]} \BibitemShut {NoStop}%
\bibitem [{\citenamefont {Candes}\ and\ \citenamefont
  {Tao}(2005)}]{candes_decoding_2005}%
  \BibitemOpen
  \bibfield  {author} {\bibinfo {author} {\bibfnamefont {E.}~\bibnamefont
  {Candes}}\ and\ \bibinfo {author} {\bibfnamefont {T.}~\bibnamefont {Tao}},\
  }\href {\doibase 10.1109/TIT.2005.858979} {\bibfield  {journal} {\bibinfo
  {journal} {IEEE Transactions on Information Theory}\ }\textbf {\bibinfo
  {volume} {51}},\ \bibinfo {pages} {4203} (\bibinfo {year}
  {2005})}\BibitemShut {NoStop}%
\bibitem [{\citenamefont {{Van den Nest}}(2013)}]{van_den_nest_universal_2013}%
  \BibitemOpen
  \bibfield  {author} {\bibinfo {author} {\bibfnamefont {M.}~\bibnamefont {{Van
  den Nest}}},\ }\href {\doibase 10.1103/PhysRevLett.110.060504} {\bibfield
  {journal} {\bibinfo  {journal} {Phys. Rev. Lett.}\ }\textbf {\bibinfo
  {volume} {110}},\ \bibinfo {pages} {060504} (\bibinfo {year}
  {2013})}\BibitemShut {NoStop}%
\bibitem [{\citenamefont {Campbell}\ and\ \citenamefont
  {O'Gorman}(2016)}]{campbell_efficient_2016}%
  \BibitemOpen
  \bibfield  {author} {\bibinfo {author} {\bibfnamefont {E.~T.}\ \bibnamefont
  {Campbell}}\ and\ \bibinfo {author} {\bibfnamefont {J.}~\bibnamefont
  {O'Gorman}},\ }\href {\doibase 10.1088/2058-9565/1/1/015007} {\bibfield
  {journal} {\bibinfo  {journal} {Quantum Science and Technology}\ }\textbf
  {\bibinfo {volume} {1}},\ \bibinfo {pages} {015007} (\bibinfo {year}
  {2016})}\BibitemShut {NoStop}%
\end{thebibliography}%

\section{Appendix}

\subsection{Robustness of magic}\label{sub:Robustness of magic}

\begin{figure}[h!]
\includegraphics[scale=0.65]{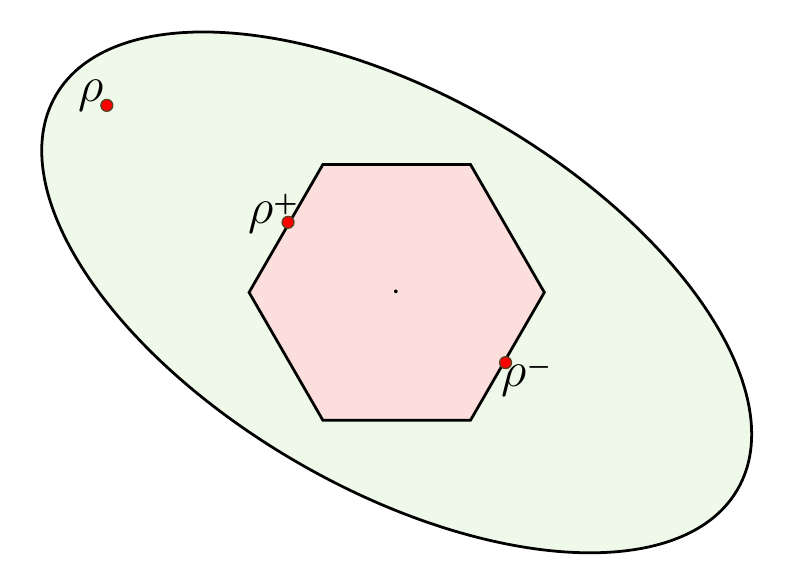}
\caption{\label{fig:geometry} Robustness of magic (geometric): The hexagon represents all possible mixtures of stabilizer states i.e., the stabilizer polytope.  Highlighted points correspond to the state of interest, $\rho$, and two stabilizer states, $\rho^{\pm}$, such that $\rho=(p+1)\rho^+-p\rho^-$. Calculating robustness of magic $\mathcal{R}(\rho)$ requires optimizing over all such decompositions but this may be recast as a linear program (Eq.~(2) of main text). This is completely analagous to the original formulation of robustness of entanglement \cite{vidal_robustness_1999} except here the extreme points of the set of free states form a discrete set. }
\end{figure}

Here we give a more extended discussion of the properties of RoM and also introduce techniques for obtaining lower bounds.  We first tackle the basic properties, which are standard for robustness measures of a resource and our proofs will parallel work elsewhere~\cite{vidal_robustness_1999}.  In contrast, our results on lower bounds (see Sec.~\ref{sec:lowerbounds}) are tailored to the magic state setting and do not have counterparts in other resource theories.  Note that robustness is mostly described here in algebraic terms but a geometric picture as in Fig \ref{fig:geometry} is natural.

\subsubsection{Basic properties}

We list the basic properties of the robustness of magic
\begin{enumerate}
    \item[R1] Faithfulness: $\mathcal{R}( \rho )=1$ if and only if $\rho$ is a stabilizer state, and otherwise $\mathcal{R}( \rho )>1$.
    \item[R2] Submultiplicativity: $\mathcal{R}( \rho_1 \otimes \rho_2  ) \leq \mathcal{R}( \rho_1 ) \cdot  \mathcal{R}(\rho_2  ) $;
    \item[R3] Monotonicity: for all trace-preserving stabilizer channels $\mathcal{E}$, we have $\mathcal{R}( \mathcal{E}(\rho) ) \leq \mathcal{R}( \rho )$.
    \item[R4] Convexity: 
$\mathcal{R}( \sum_k p_k \rho_k ) \leq \sum_k |p_k| \mathcal{R}(\rho_k)  $.
\end{enumerate}
Let us prove these properties one by one.

\noindent\textit{Faithfulness.-} If $\rho$ is a mixed-stabilizer state then by definition there exists a decomposition $\rho = \sum_i p_i \sigma_i$ where $p_i$ are positive and so setting $x_i=p_i$ we have $\sum_i |x_i|=\sum_i x_i = 1$.  Conversely, if $\rho$ is nonstabilizer state, then there is at least one negative value of $x_i$.  Furthermore, $\Tr(\rho)=1$ entails $\sum_i x_i=1$ and so we can express the absolute sum as $\sum_i |x_i| = 1 + 2 \sum_{i :  x_i<0} | x_i|$ where the new summation is only over the negative values.  We see that if a single value is negative, then the RoM exceeds unity.

\noindent\textit{Submutliplicativity.-}  Let $\rho_1$ and $\rho_2$ have optimal pseudomixtures
\begin{align}
\label{eq:opt}
	\rho_1 & =\sum_i x_i \sigma_i ,\\ \nonumber
	\rho_2 & = \sum_j y_j \sigma_j ,
\end{align}
where optimality entails $\mathcal{R}(\rho_1)=\sum_i |x_i|$ and $\mathcal{R}(\rho_2)=\sum_j |y_j|$.  We consider the composite state
\begin{equation}
	\rho_1 \otimes \rho_2 = \sum_{i,j} x_i y_j (\sigma_i \otimes \sigma_j). 	
\end{equation}
Since $\sigma_i \otimes \sigma_j$ are pure stabilizer states, this is a stabilizer pseudomixture with quasiprobabilities  $x_i y_j$.  Taking the absolute sum we have 
\begin{align}
		\sum_{i,j} |x_i y_j | & = \left( \sum_{i} |x_i| \right) \left( \sum_{j} |y_j| \right) \\ \nonumber
	& = \mathcal{R}(\rho_1)\mathcal{R}(\rho_2).
\end{align}
Therefore, $\mathcal{R}(\rho_1 \otimes \rho_2) \leq \mathcal{R}(\rho_1)\mathcal{R}(\rho_2)$.

\noindent\textit{Monotonicity.-} If $\mathcal{E}$ is stabilizer operation, then $\mathcal{E}(\sigma_i)=\sum_{j} p_{i,j} \sigma_j$ where $p_{i,j} \geq 0$.  Furthermore, if $\mathcal{E}$ is trace-preserving  then, for all $i$, we have $\sum_{j} p_{i,j}= 1$.  Combined with positivity we know  $\sum_{j} |p_{i,j}| = 1$.  Therefore, applying $\mathcal{E}$ to an arbitrary state
\begin{align}
	\mathcal{E} ( \rho ) & = \sum_{i} x_i  \sum_{j} p_{i,j}  \sigma_j,	\\ \nonumber
& = \sum_{j}  \left( \sum_{i} x_i p_{i,j} \right)  \sigma_j,	\\ \nonumber
\end{align}
which is a stabilizer pseudomixture with quasiprobabilities $x'_j=\sum_{i} x_i p_{i,j}$.  Therefore,
 \begin{align}
	\mathcal{R} \left(\mathcal{E} \left(  \rho \right) \right)& \leq  \sum_{j} |\sum_i x_i p_{i,j} | .
\end{align}
Using the triangle inequality $|\sum_k a_k| \leq  \sum_k |a_k|$ and $|ab|=|a|\cdot|b|$, we have 
\begin{align}
	\mathcal{R} \left(\mathcal{E} \left(  \rho \right) \right) & \leq \sum_i |x_i|  \sum_{j} |p_{i,j} | \\ \nonumber
& = \sum_i |x_i| \\ \nonumber
& = \mathcal{R} \left( \rho \right) . 
\end{align}
By similar methods we can show that the average robustness is non-increasing under a trace non-increasing map e.g. postselection on a particular outcome of a stabilizer POVM. The proof proceeds along the same lines as that provided in \cite{vidal_robustness_1999}.

\noindent\textit{Convexity.-} Consider a set of quantum states with optimal stabilizer pseudomixtures 
\begin{equation}
		\rho_k = \sum_i x_{i}^{(k)} \sigma_i,
\end{equation}
so that $\mathcal{R}( \rho_k  )=\sum_i | x_{i}^{(k)} |$. It follows that
\begin{equation}
	\sum_k p_k \rho_k = \sum_i \sum_k p_k x_i^{(k)} \sigma_i .
\end{equation}
Taking $x_i' = \sum_k p_k x_i^{(k)}$ as quasiprobabilities we have
\begin{equation}
	\mathcal{R} \left( \sum_k p_k \rho_k \right) \leq \sum_i |\sum_k p_k x_i^{(k)}| .
\end{equation}
Applying the triangle inequality, we have
\begin{align}
	\mathcal{R} \left( \sum_k p_k \rho_k \right) & \leq \sum_i \sum_k |p_k|\cdot | x_i^{(k)}| \\ \nonumber
& = \sum_k |p_k| \mathcal{R}( \rho_k  ).
\end{align}
This proves convexity.

Taking the logarithm of the robustness creates a related measure $LR$ such that $LR(\rho)=\log_2 ( \mathcal{R}(\rho))$.  It immediately follows that $LR$ is also monotonically decreasing under stabilizer channels.  Submultiplicativity translates into subadditivity.  Furthermore, $LR$ is faithful in the sense that $LR(\rho)=0$ if and only if $\rho$ is a stabilizer state.

\subsection{Lower bounds on RoM}\label{subsec:Lower bounds on RoM}
\label{sec:lowerbounds}

This section introduces a new quantity that we use to establish lower bounds on RoM. Such bounds are valuable since numerical methods are limited to modest numbers of qubits, whereas the lower bounds will hold for any number of qubits.

We define $\mathcal{D}$ (also called the st-norm $|| \cdots ||_{\mathrm{st}}$ in Ref.~\cite{campbell_catalysis_2011}) as
\begin{align}
  \mathcal{D} (\rho) & = \frac{1}{2^n} \sum_{ P \in \Pplus } | \tr (P \rho) | ,
\end{align}
where $\Pplus$ is the set of Pauli operators with $+1$ phase, including the identity. 
It has some useful properties, 
\begin{enumerate}
	\item[D1\phantom{*}] Convexity: $\mathcal{D}(\sum_k p_k \rho_k ) \leq \sum_k  |p_k| \mathcal{D}(\rho_k ) $ ,
	\item[D2\phantom{*}] Magic witness: if $\mathcal{D}(\rho)>1$  then $\rho$ is a nonstabilizer state,
    \item[D3\phantom{*}]  Lower bound: $\mathcal{D} (\rho) \leq \mathcal{R}(\rho)$, 
    \item[D3*]  Tighter lower bound: If $\rho$ is an $n$-qubit state then \\
\begin{equation}
	\frac{\mathcal{D}(\rho)-\frac{1}{2^n}}{\left(1 -  \frac{1}{2^n} \right)}   \leq \mathcal{R}(\rho) , \label{eqn:tighterLB}
\end{equation}
    \item[D4] Multiplicativity: $\mathcal{D}(\rho_1 \otimes \rho_2) = \mathcal{D}(\rho_1) \cdot \mathcal{D}( \rho_2)$ .
\end{enumerate}

\noindent\textit{Convexity.-}  We have by linearity of the trace that
\begin{align}
	\mathcal{D}(\rho)  &= \frac{1}{2^n} \sum_{P \in \Pplus} | \tr[ P \sum_k p_k \rho_k ]| 	\\ \nonumber
&=\frac{1}{2^n}   \sum_{P \in \Pplus} |  \sum_k  \tr[ P p_k \rho_k ]|.
\end{align}
The triangle inequality $|\sum_k a_k|\leq \sum_k |a_k|$ entails
\begin{align}
	\mathcal{D}(\rho)  &\leq  \frac{1}{2^n} \sum_{P \in \Pplus} \sum_k | \tr[ P p_k \rho_k ]|.
\end{align}
A factor $|p_k|$ can come outside the absolute value sign, and reordering the summations we have
\begin{align}
	\mathcal{D}(\rho)  &\leq \frac{1}{2^n} \sum_k | p_k |  \sum_{P \in \Pplus}  | \tr[ P \rho_k ]|  \\ \nonumber
& = \sum_k |p_k| \mathcal{D}(\rho_k) ,
\end{align}
which proves convexity.  

\noindent\textit{Magic witness.-} A pure $n$-qubit stabilizer state $\sigma_i$ has a stabilizer group $G_{\sigma_i} \subset \mathcal{P}$ (the Pauli group on $n$ qubits) containing $2^n$ elements, so that
\begin{equation}
	\sigma_i = \frac{1}{2^n} \sum_{g \in G_{\sigma_i} }	g.
\end{equation}
and such a pure state must have $\mathcal{D}(\sigma_i)=1$.  For a mixed stabilizer state, $\rho  =\sum_i p_i \sigma_i$ with $p_i>0$, and we can apply convexity so that
\begin{align}
		\mathcal{D}(\rho)  & \leq  \sum_i p_i \mathcal{D}(\sigma_i) \\ \nonumber
		 & = \sum_i p_i = 1 .
\end{align}
Since all stabilizer states have $\mathcal{D}(\rho) \leq 1$, this implies that if $\mathcal{D}(\rho)>1$ then $\rho$ must be a nonstabilizer state.  Note that some mixed stabilizer states will have $\mathcal{D}(\rho) < 1$.

\noindent\textit{Lower bound.-} Let $\rho = \sum_i x_i \sigma_i$ be an optimal stabilizer pseudomixture.  By convexity we have
\begin{align}
	\mathcal{D}(\rho) \leq \sum_i|x_i| \mathcal{D}(\sigma_i),
\end{align}
and using $\mathcal{D}(\sigma_i)=1$ we have
\begin{align}
	\mathcal{D}(\rho) \leq \sum_i|x_i|  = \mathcal{R}(\rho),
\end{align}
which completes the proof.

\noindent\textit{Tighter lower bound.-} Next, we prove an even tighter lower bound.  Asymptotically, the tighter bound is identical to the above lower bound, but it is much tighter for modest numbers of qubits.  We begin by revisiting the first line of the convexity proof and split the sum over $\Pplus$ into one term for $\id \in \Pplus$   and the remainder $\Pstar=\Pplus \setminus \id$ so that
\begin{align}
\label{Drho}
	\mathcal{D}(\rho)  &= \frac{1}{2^n} \left(|\tr[ \rho ] | + \sum_{P \in \Pstar} |\tr[ P \rho ] |  \right) \\ \nonumber
	 &= \frac{1}{2^n} \left(1 + \sum_{P \in \Pstar} |\tr[ P \rho ] |  \right) 
\end{align}
Writing $\rho$ as its optimal stabilizer pseudomixture and applying the triangle inequality we arrive at
\begin{align}
	\mathcal{D}(\rho)  & \leq \frac{1}{2^n} \left(1 + \sum_i |x_i|\sum_{P \in \Pstar}  |\tr[ P \sigma_i ] |  \right) .
\end{align}
We note that
\begin{align}
	\frac{1}{2^n}	\sum_{P \in \Pstar}  |\tr[ P \sigma_i ] | & = \frac{1}{2^n}\left( \sum_{P \in \Pplus}  |\tr[ P \sigma_i ] | \right) - \frac{1}{2^n} \\ \nonumber
& = \mathcal{D}(\sigma_i) - \frac{1}{2^n} \\ \nonumber
& = 1 - \frac{1}{2^n} .
\end{align}
Applying this to Eq.~\eqref{Drho} gives
\begin{align}
	\mathcal{D}(\rho)  & \leq \frac{1}{2^n} + \left( \sum_i |x_i| \right) \left(1 -  \frac{1}{2^n} \right) .
\end{align}
Using $\mathcal{R}(\rho)=\sum_i |x_i|$ and rearranging for $\mathcal{R}(\rho)$, we find the tighter lower bound stated earlier.

\noindent\textit{Multiplicativity.-} For a product state we have
\begin{equation}
	\mathcal{D}(\rho_1 \otimes \rho_2 ) =\frac{1}{2^n} \sum_{\mathcal{P \in \Pplus }} | \tr[ P (\rho_1 \otimes \rho_2) ] |	.
\end{equation}
Every $P \in \Pplus$ can be written as a product $P = P_1 \otimes P_2$ where $P_1 \in \Pplus$ and $P_2 \in \Pplus$ for the smaller Hilbert spaces. Therefore
\begin{equation}
	\mathcal{D}(\rho_1 \otimes \rho_2 ) = \frac{1}{2^n}  \sum_{P_1, P_2 \in \Pplus }  | \tr[ (P_1 \rho_1) \otimes ( P_2 \rho_2) ] |	.
\end{equation}
The trace is multiplicative with respect to tensor products ($\tr[A\otimes B] =\tr[A] \tr [B]$) and so
\begin{align}
	\mathcal{D}(\rho_1 \otimes \rho_2 ) & = \frac{1}{2^n}  \sum_{P_1, P_2 \in \Pplus } | \tr[ (P_1 \rho_1) |    \cdot | \tr[  ( P_2 \rho_2) ] |	\\ \nonumber
& =   \sum_{P_1, P_2 \in \Pplus } \frac{1}{2^{n_1}} | \tr[ (P_1 \rho_1) |    \cdot  \frac{1}{2^{n_2}}| \tr[  ( P_2 \rho_2) ] |	\\ \nonumber
 & = \mathcal{D}(\rho_1) \mathcal{D}(\rho_2),
\end{align}
where in the second line we split $n=n_1 +n_2$ where $n_1$ and $n_2$ are the number of qubits in systems 1 and 2.  This proves multiplicativity.

The combination of multiplicativity with the lower bound property entails that $\mathcal{D}(\rho)^n = \mathcal{D}(\rho^{\otimes n}) \leq \mathcal{R}(\rho^{\otimes n})$.  This enables us to lower bound robustness for large $n$ even though numerically finding the robustness is difficult for large $n$.  Let us apply these techniques to mixed versions of the magic states singled out by Bravyi and Kitaev \cite{bravyi_universal_2005} (we replace their $T$ with $F$ to avoid notational confusion),
\begin{align}
	\rho_H(r) & = \frac{1}{2}\left( \id + r \frac{X+Z}{\sqrt{2}} \right) ,\\ 
	\rho_F(r) & = \frac{1}{2}\left( \id + r \frac{X+Y+Z}{\sqrt{3}} \right) \label{eqn:FaceState},
\end{align}
where $r=\pm 1$ for pure states and $-1 < r < +1$ for mixed states.  It is easy to verify that 
\begin{align}
    \mathcal{D}(\rho_H(r)) & = \frac{1}{2}(1+\sqrt{2} r) \label{eqn:DofH}  ,\\
    \mathcal{D}(\rho_F(r)) & =\frac{1}{2}(1+\sqrt{3} r)  .
\end{align}
Therefore, for pure states $r=1$, we can conclude
\begin{align}
    1.207^n & \leq \mathcal{R}(\rho_H(1)^{\otimes n}) , \\
    1.366^n & \leq \mathcal{R}(\rho_F(1)^{\otimes n})  .
\end{align}
Using multiplicativity with the tighter lower bound, for a $m$-qubit state $\rho$ we have 
\begin{align}
	\frac{\mathcal{D}(\rho)^n-\frac{1} {2^{nm}} } {\left(1 -  \frac{1}{2^{nm}} \right) }   \leq \mathcal{R}(\rho^{\otimes n}) .
\end{align}
Whenever $\mathcal{D}(\rho)>1$, the lower bound clearly approaches $\mathcal{D}(\rho)^n$ as $n\rightarrow \infty$.

\subsection{Robustness of Particular States}\label{sub:Robustness of Particular States}

\subsubsection{Robustness of States from the Third Level of the Clifford Hierarchy}
In the main text we gave the numerical value for the robustness of $t$-fold copies of $\ket{H}$. Here we provide the exact symbolic expressions for small $t$,
\begin{align}
\mathcal{R}(\ket{H^{}})&=\sqrt{2}\approx 1.4142, \\
\mathcal{R}(\ket{H^{\otimes 2}})&=\frac{1+3\sqrt{2}}{3}\approx 1.7476, \\
\mathcal{R}(\ket{H^{\otimes 3}})&=\frac{1+4\sqrt{2}}{3}\approx 2.2190, \\
\mathcal{R}(\ket{H^{\otimes 4}})&=\frac{3+8\sqrt{2}}{5}\approx 2.8627. 
\end{align}
We also have
\begin{align}
\mathcal{R}(\ket{H^{\otimes 5}})&\approx 3.68705 
\end{align}
although without a neat symbolic expression.

For $t$ up to 5 we have calculated expressions for the exact value of $\mathcal{R}(\ket{H^{\otimes t}})$. For $t>5$ we can also place bounds on this quantity as shown in Table \ref{Tab:bounds_on_Rob}. In addition, we can give a full classification of all 3-qubit diagonal gates from the set generated by CNOT$+T$ as in Table~\ref{Tab:three_qubit_classification}. A similar table can be constructed for 4-qubit diagonal gates using the data provided in Supplementary Material.

\begin{table*}[t]
\centering
\begin{tabular}{rcccc}
\hline
 &\hspace{0.25cm} Lower from $\mathcal{D}$ \eqref{eqn:tighterLB} and \eqref{eqn:DofH} \hspace{0.25cm} &  Lower from $\mathcal{R}(\ket{U})$  & $U\in\langle \textsc{CNOT}, T\rangle$ & Upper  \\
\hline
$\mathcal{R}(\ket{H^{\otimes 6}})$ \quad & 3.1269 & 4.7031 & $T_1CS_{24,35,45}$ & 4.9238\\
$\mathcal{R}(\ket{H^{\otimes 7}})$ \quad  & 3.75592 & 5.5242 & $T_4CS_{13,24,35,45}$ & 6.3523  \\
$\mathcal{R}(\ket{H^{\otimes 8}})$ \quad  & 4.52157 & 5.7934 & $T_5CS_{12,35,45}$ & 8.1953 \\
$\mathcal{R}(\ket{H^{\otimes 9}})$ \quad  & 5.4501 & 6.2625 & $T_{1,2,3}CS_{12,14,25}CCZ_{345}$            & 10.555 \\
$\mathcal{R}(\ket{H^{\otimes 10}})$ \quad  & 6.5738 & 6.8995 & $T_{1}CS_{12,23,14,25,45}CCZ_{345}$              & 13.594 \\
$\mathcal{R}(\ket{H^{\otimes 11}})$ \quad  & 7.9321 & 6.9255 & $T_5CS_{23,24,45}CCZ_{125,345}$              & 17.341 \\
\hline
\end{tabular}
\caption{\label{Tab:bounds_on_Rob} Bounds on $\mathcal{R}(\ket{H^{\otimes t}})$ for $6\leq t \leq 11$: The upper bound is given by $\mathcal{R}(\ket{H^{\otimes t}})\leq \mathcal{R}(\ket{H^{\otimes \left \lfloor{\tfrac{t}{2}}\right \rfloor}})\mathcal{R}(\ket{H^{\otimes \left \lceil{\tfrac{t}{2}}\right \rceil}})$ which holds because of submultiplicativity of robustness. The first lower bound derives from expression \eqref{eqn:tighterLB} using $\mathcal{D}(\ket{H^{\otimes t}})=\left[\mathcal{D}(\ket{H})\right]^{ t}=\left[\frac{1+\sqrt{2}}{2}\right]^t$ \eqref{eqn:DofH}. The second lower bound corresponds to $\mathcal{R}(\ket{U})$ where $U$, provided in the column headed $U\in\langle \textsc{CNOT}, T\rangle$, is synthesizable over the CNOT$+T$ gate set using $t$ $T$ gates~\cite{amy_t-count_2016}. Clearly any such $U$ satisfies $\mathcal{R}(\ket{U}) \leq \mathcal{R}(\ket{H^{\otimes t}}) $. Another way of interpreting this column is to note that these $U$ are the most robust gates we know how to construct over the CNOT$+T$ gate set using $t\in\{6,7,\ldots,11\}$ $T$ gates. There are no five-qubit $U\in\langle \textsc{CNOT}, T\rangle$ that require more than 11 $T$ gates.}
\end{table*}

\begin{table}[h]
\centering
\begin{tabular}{rcccc}
\hline
$\mathcal{R}(\ket{U})$ & $T$ cost & $U$   & $T$ cost$'$  & $U'$  \\
\hline
1.41421 \quad  & 1 &$T_1$ &  &  \\
1.74755 \quad  & 2 &$T_{1,2}$&  &   \\
2.2     \quad  & 3 &$CS_{12}$&  & \\
2.21895 \quad  & 3 &$T_{1,2,3}$&     &         \\
2.55556 \quad  & 4 &$CS_{12,13}$ & 7  &  $CCZ_{123}$           \\
2.80061 \quad  & 4 &$T_1CS_{23}$&     &            \\
3.12132 \quad  & 5 &$T_1CS_{12,13}$ & 6  &  $T_1CCZ_{123}$          \\
\hline
\end{tabular}
\caption{\label{Tab:three_qubit_classification} Three qubit classification: All the three-qubit diagonal gates from the third level of the Clifford hierarchy can be classified by the robustness of the associated resource state $\ket{U}=U\ket{+}$. The cost of synthesizing these gates over the Clifford$+T$ gate set is provided. Two gates $U$ and $U'$ with the same robustness and different $T$ cost are equivalent via the construction described in connection with Eq.~(7) of the main text.}
\end{table}

\subsubsection{Numerical maximization of Robustness}

\begin{figure}[h!]
\includegraphics[scale=0.95]{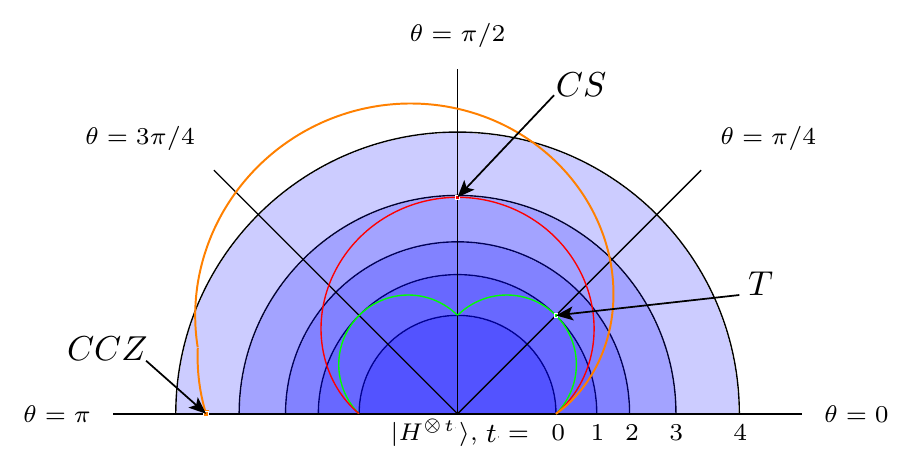}
\caption{\label{fig:concentric} Polar plot where the radial coordinate denotes robustness and the angular component represents the last entry in a $m$-qubit resource state $(1,\ldots,e^{i\theta})^T/2^{m/2}$. Concentric semi-circles correspond to robustness $\mathcal{R}(\ket{H^{\otimes t}})$ for $t=0,1,\ldots,4$. Highlighted points correspond to resource states for $T$, $CS$ and $CCZ$. We see that $\mathcal{R}(\ket{H^{\otimes 3}})$ is only slightly greater than $\mathcal{R}(\ket{CS})$, meaning that $CS$ synthesis using 3 $T$ gates as in Fig.~1(c) of main text is optimal and no ancilla-assisted or non-deterministic strategy could be more parsimonious.}
\end{figure}

Given that we have an operationally relevant quantifier of magic, it is interesting to consider for which states $\rho$ the quantity $\mathcal{R}(\rho)$ is maximized. For a single qubit the state $\ket{F}$~\cite{bravyi_universal_2005} with Bloch vector $(x,y,z)=(1,1,1)/\sqrt{3}$ given in \eqref{eqn:FaceState} has maximal robustness $\sqrt{3}$. Multiple copies of this state have robustness
\begin{align}
\mathcal{R}(\ket{F^{\otimes 2}})&=\frac{1+2\sqrt{3}}{2}\approx 2.232, \\
\mathcal{R}(\ket{F^{\otimes 3}})&=\frac{1+3\sqrt{3}}{2}\approx 3.098, \\
\mathcal{R}(\ket{F^{\otimes 4}})&=\frac{13+20\sqrt{3}}{11}\approx 4.331. 
\end{align}
For two qubits, robustness is maximized at $\sqrt{5}$ by a state that is maximally outside one of the facets of the 2-qubit stabilizer polytope (Table 2 Line 5 of~\cite{reichardt_quantum_2009}). The state $(1,1,1,i)/2$ is maximally robust amongst all ``flat'' (equally-weighted) 2-qubit states achieving $\mathcal{R}=2.2$. For 3 qubits the most robust state at $\mathcal{R}=3.8$ is the so-called Hoggar state~\cite{hoggar_64_1998},
\begin{align}
\ket{\text{Hoggar}}=\frac{1}{\sqrt{6}}\left(\begin{array}{c}
1+i \\ 
0 \\ 
-1 \\ 
1 \\ 
-i \\ 
1 \\ 
0 \\ 
0
\end{array}  \right),
\end{align}
 a fiducial vector for a symmetric, informationally complete positive operator valued measure (SIC-POVM) covariant under the 3-qubit Pauli group. Some of these states were already classified as highly non-stabilizer by other means~\cite{andersson_states_2015}. In Fig.~\ref{fig:concentric} we depict the robustness of a one-parameter family of $m$-qubit states, for $m\in\{1,2,3\}$. From this figure we can read off $\mathcal{R}(\ket{H^{\otimes 2}})<\mathcal{R}(\ket{CS})<\mathcal{R}(\ket{H^{\otimes 3}})$ and 
 $\mathcal{R}(\ket{H^{\otimes 3}})<\mathcal{R}(\ket{CCZ})<\mathcal{R}(\ket{H^{\otimes 4}})$. We can also see that a doubly-controlled rotation of angle $\theta=\pi/2$---a $CCS$ gate---would require at least $5$ $T$ gates to implement.

By exploiting the \Jam isomorphism, as in e.g.~\cite{harrow_robustness_2003,van_dam_noise_2011}, we can also investigate the robustness of operations. In particular, for every unitary $U \in \textrm{U}(2^n)$ we can associate a state $\ket{J_U}=(\id\otimes U)\sum_{j\in \mathbb{Z}_2^n}\ket{j,j}$. Maximizing the quantity $\mathcal{R}(\ket{J_U})$ tells us the most robust unitary operations. For a single qubit we numerically find the optimal unitary to be
\begin{align}
U&=\frac{1}{\sqrt{2}}\left(\begin{array}{cc}
1 & e^{i \frac{3 \pi}{4}} \\ 
e^{-i \frac{\pi}{4}} & 1
\end{array}\right),\\
\mathcal{R}(\ket{J_U})&=\mathcal{R}(\ket{H^{\otimes 2}})=\frac{1+3\sqrt{2}}{3},
\end{align}
whereas optimizing over two-qubit unitaries we find
\begin{align}
U&=\frac{1}{5}\left(\begin{array}{cccc}
-1-2i & 3+i & 1-3i & 0 \\ 
1-3i & -3-i & -1-2i & 0\\ 
3+i & 1+2i & -3-i  & 0\\ 
0 & 0 & 0 & 5
\end{array}\right),\\
\mathcal{R}(\ket{J_U})&=\frac{23}{5}.
\end{align}

\subsection{Interconvertability}\label{sub:Interconvertability}

Here we present some further examples of $T$-saving synthesis of diagonal gates from the CNOT$+T$ gate set. The savings come from Clifford equivalence of magic states as discussed in the main text in the context of equation $(7)$;
\begin{align}
T_{1,2}CCZ_{345}  &\overset{8 \rightarrow 6}{\longrightarrow}T_{1,2}CS_{35,45}, \\
T_{1,2,5}CCZ_{345}  &\overset{8 \rightarrow 7}{\longrightarrow}T_{1,2,5}CS_{35,45} ,\\
T_{1,2,3,4}CS_{23,24,34}  &\overset{7 \rightarrow 6}{\longrightarrow} T_{1,4}CS_{24,34}, \\
CS_{12}CCZ_{345} &\overset{9 \rightarrow 7}{\longrightarrow}CS_{12,35,45},\\
T_{5}CS_{12,25}CCZ_{345} &\overset{9 \rightarrow 7}{\longrightarrow}T_{5}CS_{14,25,35,45}.
\end{align}
These are all provably optimal in the sense that the $T$ cost on the right hand side is the minimum possible, as it is the smallest integer $t$ satisfying $\mathcal{R}(\ket{U}) \leq \mathcal{R}(\ket{H^{\otimes t}}) $.

We also make a brief comment on resource interconvertability. Are there optimal transformations that convert $n$-qubit $\ket{\phi}$ to $m$-qubit $\ket{\psi}$ ,  $m\neq n$, with $\mathcal{R}(\ket{\phi})=\mathcal{R}(\ket{\psi})$? At least one such transformation does exist. In the range $ 0 \leq \theta \leq \arctan \frac{1}{3}$, two copies of the equatorial state $\ket{\phi}=(\ket{0}+e^{i\phi}\ket{1})/\sqrt{2}$ have robustness 
\begin{align*}
\mathcal{R}(\ket{\phi^{\otimes 2}}) =(2 \sin \phi+\sin 2 \phi +\cos 2 \phi +1)/2
\end{align*}
but a measurement of Pauli operator $ZY$ on $\ket{\phi^{\otimes 2}}$, followed by a CNOT, creates the state $\ket{\psi\pm}\ket{Y\pm}$ depending on the $ZY$ outcome, where $ \ket{\psi\pm}$ has Bloch vector $\vec{r}=(\cos ^2(\theta ),\sin (\theta ) \cos ( \theta ),\pm \sin (\theta ))$ hence $||\vec{r}||_1=\mathcal{R}(\ket{\psi\pm})=\mathcal{R}(\ket{\phi^{\otimes 2}})$. Finding further examples of such transformations could find application in magic state preparation. The foregoing leads naturally to the question of asymptotic interconvertibility, much studied in the context of entanglement, where one seeks to maximize the rate $m/n$ in the limit $n\rightarrow \infty$ of a stabilizer operation mapping $\ket{\phi^{\otimes n}} \rightarrow \ket{\psi^{\otimes m}}$. In this work we have dealt solely with deterministic gate synthesis but the extension to probabilistic (depending on measurement outcomes) circuit synthesis is natural; the expected robustness $\mathbb{E}[\mathcal{R}(\ket{\phi})]$ is now the relevant quantity to compare to e.g, $\mathcal{R}(\ket{H^{\otimes n}})$. Our choice to focus on synthesizing gates from the third level of the Clifford hierarchy was motivated by the fact that any measurement outcomes $m_i$ in Fig.~(1) of main text can be corrected for with a Clifford gate. Other resource states will not have this property and unwanted measurement outcomes could result in either disposing of the output state or correcting with a non-Clifford gate whose cost would also need to be taken into account.

%\bibliography{Robustness_appendix}

\end{document}